\documentclass[A4paper,useAMS,usenatbib]{mn2e}
\usepackage{graphicx,amssymb,color}
\usepackage{lastpage}
\usepackage[fleqn]{amsmath}
\usepackage[normalem]{ulem}
\input psfig

\title[Sublimation of planetesimals in white dwarf systems]
{Sublimation-induced orbital perturbations of extrasolar active asteroids and comets:
application to white dwarf systems}
\author[Veras, Eggl \& G\"{a}nsicke]{
Dimitri Veras$^{1}$\thanks{E-mail:d.veras@warwick.ac.uk},
Siegfried Eggl$^{2}$,
Boris T. G\"{a}nsicke$^{1}$
\\
$^{1}$Department of Physics, University of Warwick, Coventry CV4 7AL, UK
\\
$^{2}$IMCCE Observatroire de Paris, Univ. Lille 1, UPMC, 77 Av. Denfert-Rochereau, 75014 Paris, France
}

\begin{document}

\date{Accepted 2015 June 23. Received 2015 June 23; in original form 2015 March 26}

\pagerange{\pageref{firstpage}--\pageref{lastpage}} \pubyear{XXXX} 

\maketitle

\label{firstpage}

\begin{abstract}
The metal budgets in some white dwarf (WD) atmospheres reveal that volatile-rich circumstellar bodies must both exist in extrasolar systems and survive the giant branch phases of stellar evolution.  The resulting behaviour of these active asteroids or comets which orbit WDs is not well-understood, but may be be strongly influenced by sublimation due to stellar radiation.  Here we develop a model, generally applicable to any extrasolar system with a main sequence or WD star, that traces sublimation-induced orbital element changes in approximately km-sized extrasolar minor planets and comets traveling within hundreds of au.  We derive evolution equations on orbital timescales and for arbitrarily steep power-law sublimation dependencies on distance, and place our model in a Solar system context.  We also demonstrate the importance of coupling sublimation and general relativity, and the orbital consequences of outgassing in arbitrary directions.  We prove that nongravitational accelerations alone cannot result in orbit crossing with the WD disruption radius, but may shrink or expand the orbit by up to several au after a single pericentre passage, potentially affecting subsequent interactions with remnant debris and planets.  Our analysis suggests that extant planets must exist in polluted WD systems.
\end{abstract}


\section{Introduction}

Water is known to exist in Solar system planets, moons, comets and 
asteroids.  Even after almost 5 Gyr of Solar evolution, 
active dynamical processes involving water, like the ventilation of plumes 
on Enceladus \citep{hanetal2006}, 
persist.  Comets still routinely shed their volatile content during close 
approaches to the Sun.  The recent discovery of active asteroids 
\citep[see][for a review]{jewetal2015} demonstrate that significant amounts 
of ice can exist in these bodies.  But what will
happen to the ice and water when the Sun turns off of the main sequence?

The giant branch (GB) phases of evolution are violent, featuring stellar mass loss and intense radiation.
 In principle, this radiation will evaporate any surface water content of orbiting bodies out to several au,
although internal volatiles may be retained \citep{jurxu2010,jurxu2012}.  
In this paper, we refer to these orbiting bodies as two types: \textit{comets} and \textit{minor planets}. In the context of exoplanetary systems, we define comets as bodies that formed 
beyond the snow/frost line, the line at which water and other volatiles froze out
during system formation.  The volatile content of comets may approach 100 per cent, and represent a 
complex cocktail 
of e.g. water ice, carbon dioxide, carbon monoxide, and methane.  Alternatively, minor planets
formed closer to the star, and hence are less volatile-rich.

Whatever events do occur during the GB phases, we have concrete observational 
evidence of the final outcome after the stars have become white dwarfs
(WDs): atmospheric metal pollution with a high volatile content \citep{faretal2013,radetal2015}.
In the case of the WD named GD 61, \cite{faretal2013} found that the progenitor 
of the debris in the atmosphere was 26 per cent water by mass; \cite{radetal2015} later found
that the atmosphere of the WD named SDSS J124231.07$+$522626.6 is enriched with elements
that arose from a progenitor that was 38 per cent water by mass.

Between one quarter and one half of all WDs harbour metal-polluted 
atmospheres \citep{zucetal2003,zucetal2010,koeetal2014}.  The dense nature of WDs and the 
resulting fast sinking times \citep[e.g. Fig. 1 of][]{wyaetal2014} in their atmospheres indicate that the currently-observed metals do not arise from GB remnants nor internal dredge-up.  Further, the spatial distribution and
density of interstellar medium clouds prevent the metal pollution from originating there 
\citep{aanetal1993,frietal2004,jura2006,kilred2007,faretal2010}.  Instead, 
the metal pollution must arise from circumstellar material; bolstering this hypothesis is the presence of orbiting
dusty and gaseous discs.

All 37 discs discovered around WDs so far contain dust \citep{zucbec1987,becetal2005,kiletal2005,reaetal2005,faretal2009,beretal2014,rocetal2015}; 7 of them also contain strong gaseous components \citep{ganetal2006,ganetal2007,ganetal2008,gansicke2011,faretal2012,meletal2012,wiletal2014,manetal2015,wiletal2015}.  These discs exist in exceptionally compact configurations ($\sim 0.6-1.2 R_{\odot}$, where $R_{\odot}$ is the Solar radius) that are unlike anything which is observed in the Solar system, and are thought to arise from the tidal disruption of minor planets \citep{graetal1990,jura2003,debetal2012,beasok2013,veretal2014a} and subsequent circularization of the resulting debris \citep{veretal2015b}.  The disruption of comets is assumed to provide a smaller contribution to these discs, but is important otherwise for delivering large quantities of hydrogen to WD atmospheres \citep{veretal2014c,stoetal2015}.

The minor planets or comets disrupt by entering the disruption (or Roche) radius -- located at a distance of approximately $1R_{\odot}$ -- from a relatively great distance of at least several au, and hence contain eccentricities near unity.  The semimajor axes of the minor planets or comets must be several au because otherwise they would have been engulfed by the expanding stellar envelope of the GB star.  Surviving grains, pebbles, comets, and minor and major planets are affected by a variety of other forces which may perturb these bodies to greater distances \citep{veretal2015a}.  Planets are usually too large to be affected by forces other than stellar mass loss \citep{omarov1962,hadjidemetriou1963,veretal2011,adaetal2013,veretal2013a}. Minor planets and comets, however, will also be perturbed by the Yarkovsky effect \citep{veretal2015a}, the YORP effect \citep{veretal2014b}, Poynting-Robertson drag and radiation pressure \citep{bonwya2010,veretal2015a}, and possibly stellar wind drag \citep{donetal2010,veretal2015a}, much less any binary stellar companions \citep{kraper2012,vertou2012} or surviving planets in the system \citep{debsig2002,bonetal2011,debetal2012,veretal2013b,voyetal2013,frehan2014,musetal2014,vergae2015}.  

Such complications ensure that minor planets and comets will exist on a wide variety of orbits around WDs, and not necessarily ones which enter the WD's disruption sphere.  So what happens to the volatile-rich minor planets and comets which {\it miss} the disruption sphere?  In this work, we isolate the orbital effects of the sublimation and outgassing of volatiles due to WD radiation, and focus on the minor planets with highly eccentric orbits that fail to reach the WD disruption radius.  
Our results are also applicable to a broad variety of comets, as long as their volatile content is not high enough to lead to break up or non-periodic behaviour.  We create orbital models which are widely applicable to nearly any extrasolar system assuming that no major planets interfere with the minor planet's orbit.

In Section 2, we outline our model assumptions for volatile sublimation.  Sections 3 contains the resulting unaveraged orbital element evolution equations, along with a few applications, demonstrating both the interplay of general relativity and sublimation and the nearly-fixed nature of the orbital pericentre.  We then reconsider how this pericentre might change due to outgassing in Section 4, where we also relate the extrasolar and Solar system sublimation models.  We discuss the implications in Section 5 and conclude in Section 6.  The Appendix contains an extension of our sublimation model that removes the key assumption of a specific power-law exponent.

\section{Sublimation model basics and motivation} \label{SUBLIMATION}

Centuries of observations of Solar system comets reveal orbital changes
at each apparition.  These changes, which can represent shifts in semimajor
axis and perihelion of tenths of an au on short timescales
(e.g. see Section 5 and Figure 3 of \citealt*{szutowicz2000}, and 
\citealt*{maquet2015}) may have dramatic
long-term consequences.  Sungrazing comets often are destroyed 
outright \citep[e.g.][]{broetal2011}, such as
C/2012 S1 ISON \citep{knibat2014}.  Orbital alterations are 
primarily caused by two factors: (1) gravitational
interactions due to the architecture of
the Solar system, which contains many major and minor planets and 
moons, and (2)
nongravitational forces.  

The presence of volatiles ensures that the nongravitational forces from
sublimation and outgassing will dominate those from radiation pressure,
Poynting-Robertson drag, the Yarkovsky effect and YORP.  By {\it sublimation}
we refer to the mass exodus of surface particles which are distributed approximately
evenly over large parts of the minor planet or cometary surface.  
By {\it outgassing} we refer to localized violent eruptions of material in 
random directions.  This section and Section 3 covers sublimation, and Section 4 covers 
outgassing.

Minor planets which are strongly affected by sublimation or outgassing
are known as ``active asteroids'' \citep[e.g.][]{jewetal2015}.
Classically, strong outgassing as seen in some comets takes the form of a visibly 
dramatic tail.  Modern observations
reveal that the situation is much more complex, sometimes including up to six
tails for an individual object \citep{jewetal2013}.

Consequently, small-body sublimation and outgassing has predominately been investigated
observationally in the context of the Solar system and from the perspective of 
an observer on Earth.  Theoretical developments have largely followed suit.  Here, 
however, we consider sublimation from a more general perspective, for application
to exoplanetary systems.  For WD
systems, we have not yet directly detected comets, 
major planets\footnote{The object WD 0806-661 b is by many definitions
a bona-fide planet orbiting a WD at a distance of about 2500 au
\citep{luhetal2011}.  Further, an increasingly strong case is
being made for a pair of purported planets which orbit a
WD - M star binary \citep{maretal2014}.}, moons nor any 
intact minor planets, and hence cannot utilise many of the tools developed
for or assumptions about the Solar system.

\subsection{Physical properties of the minor planet} 

Denote the mass of the minor planet as $M$, which contains both a rocky 
non-volatile component of mass $M_{\rm c}$ and a volatile component
of mass $M_{\rm v}$ such that

\begin{equation}
M = M_{\rm c} + M_{\rm v}.
\end{equation}

\noindent{}The non-volatile content may be thought of as the 
minor planet's {\it core}, although we need not make a priori assumptions about 
just where the volatiles are located.
We consider only minor planets that will remain largely intact,
such that they do not fragment or sublimate away a large fraction
of their mass.  Consequently,
the assumption that we do make is that $M_{\rm c} \gg M_{\rm v}$,
so that $M \approx M_{\rm c}$ remains constant over the number of pericentre
passages needed for all of $M_{\rm v}$
to sublimate away.  This assumption is robust for minor planets and many
types of comets.
Alternately, for some comets, $M_{\rm v}/M$ may approach unity.  In this case,
the dynamical evolution can change in more drastic ways than we report here. 

The rotation state of the minor planet will crucially determine where
outgassing will occur, with consequences for the minor planet's orbital
evolution. \cite{cheyeo2005} have shown that outgassing via jets can be modeled
in an average sense, if the rotation of the object is well defined. Their model suggests 
that outgassing is predominately happening on the starlit 
hemisphere on slow rotators with low thermal inertia. 
Yet, outgassing itself will likely change the rotation
state of the minor planet \citep{motetal2014}, thereby randomizing the perturbations
and emphasizing the stochasticity of the process of outgassing.  Stochastic
variations intrinsically contain time-dependent variations in direction
and magnitude, as would occur in a tumbling minor planet.  The actual accelerations
will be probabilistically distributed \citep[e.g.][]{pierret2014}.  

Treating potentially complex rotation is beyond the scope of the present work and 
beyond the scope of our knowledge of minor planets in extrasolar systems.  Instead, 
we consider outgassing from a general perspective in Section 4, rather than assuming 
outgassing occurs only on the starlit hemisphere, as in \cite{fesetal1993}.

\subsection{Orbital change of the minor planet} 

Many forces contribute to the orbital motion of the minor planet,
including those due to radiation, sublimation and outgassing.
Further, minor planets with orbital pericentres close to the WD
might be significantly affected by general relativity (GR).  If
the minor planet has an orbital period similar to those of exo-Oort
cloud comets, then near apocentre both Galactic tides and
stellar flybys can cause significant perturbations.

Strictly, when a body (like a minor planet) loses mass anisotropically, 
its orbit is perturbed by both linear momentum recoil and mass loss.
Although fully described analytically \citep{veretal2013b}, 
the equations of motion due solely to mass loss can be
ignored here because $M \ll M_{\rm WD}$, where $M_{\rm WD}$ is the
mass of the WD.  The overall change in mass of the system
due to sublimation plus outgassing is negligible.
The primary perturbations on orbits due to sublimation and outgassing
is linear momentum recoil.

\cite{whipple1950} used conservation of momentum to show that the nongravitational
acceleration on a minor planet ($\vec{J}_{\rm NG}$) is related to its ejecta through

\begin{equation}
M \vec{J}_{\rm NG} = -D m_{\rm s} \vec{v}_{\rm g}
,
\label{whip}
\end{equation}

\noindent{}where $m_{\rm s}$ is the mass of one molecule of the
molecular species that is sublimating, $\vec{v}_{\rm g}$ is the mean ejection velocity of
that species, and $D$ is the number of molecules of that species sublimating per time.
For our numerical computations, we will assume the sublimating species is water ice (H$_2$O),
and hence will adopt $m_{\rm s} = 2.99 \times 10^{-26}$ kg.  We assume that most of the sublimation
occurs on the starlit hemisphere because the minor planet's tiny thermal inertia will
ensure a near-zero thermal wave lag angle \citep{groetal2013}.  Consequently, the complex
physics which relates thermal balance, rotation, sublimation and outgassing is simplified
here through the values of $D$ and $\vec{v}_{\rm g}$.


The equation of motion for the minor planet is

\begin{equation}
\ddot{\vec{r}} =
-\frac{G\left(M_{\rm WD} + M_{\rm c}\right)\vec{r}}{r^3}
+ 
\ddot{\vec{r}}_{\rm sub}
+
\ddot{\vec{r}}_{\rm extra}
,
\label{nongrav1}
\end{equation}

\noindent{}where

\begin{equation}
\ddot{\vec{r}}_{\rm sub} =
\Theta\left(M_{\rm v}\right)
\vec{J}_{\rm NG}
,
\label{nongrav1b}
\end{equation}

\noindent{}$\vec{r}$ is the distance between the minor planet and
the WD, $\ddot{\vec{r}}_{\rm sub}$ is the acceleration due to
sublimation, $\ddot{\vec{r}}_{\rm extra}$ is the acceleration due to other
forces, and $\Theta$ is the Heaviside step function such that 
$\Theta(M_{\rm v})=1$ when $M_{\rm v} > 0$, and zero otherwise.  
Other forces which must be included in $\ddot{\vec{r}}_{\rm extra}$
depend on the system being studied.  These additive terms from other
forces have been used or applied in previous investigations: mass loss
\citep{omarov1962,hadjidemetriou1963,veretal2013a}, the Yarkovsky effect
\citep{veretal2015a}, the YORP effect \citep{veretal2014b}, Galactic
tides \citep{heitre1986,ricfro1987,fouetal2006,vereva2013b} and stellar
flybys in the Galactic field \citep{freetal2004,zaktre2004,vermoe2012}. 

One effect which might strongly influence the orbit of any minor planet
or comet with a pericentre close to the star is GR.  Solar
system studies have indicated the importance of GR in calculations of
cometary orbital solutions \citep[e.g.][]{yeoetal1996,maqetal2012}.
Consequently, we do incorporate GR into our formalism.  We do not
consider the other effects mentioned above any further,
and hence set $\ddot{\vec{r}}_{\rm extra} = \ddot{\vec{r}}_{\rm GR}$.  
This assumption is reasonable because large minor planets with low
thermal inertia are not strongly affected by Yarkovsky forces.
Similarly, sublimation and outgassing dominate the evolution of
minor planet spin as long as volatiles are present.

We use the complete expression for the 1PN (leading 
post-Newtonian) term of GR in the
limit of a zero-mass minor planet 
\citep[equation 3.190 of][]{beutler2005}

\begin{equation}
\ddot{\vec{r}}_{\rm GR} = 
\frac{G M_{\rm WD}}{c^2 r^3}
\left[
\left(
\frac{4GM_{\rm WD}}{r}
-
v^2
\right)
\vec{r}
+
4
\left(
\vec{v}\cdot\vec{r}
\right)
\vec{v}
\right]
\end{equation}

\noindent{}where $\vec{v}$ is the velocity of the minor planet
with respect to the star and $c$ is the speed of light.

Now consider equations equations (\ref{whip})-(\ref{nongrav1b}) in greater depth.
In order to develop physical intuition for how the 2-body orbit changes
due to linear momentum recoil, we insert equation (\ref{nongrav1b}) into equation (\ref{nongrav1}) 
and reexpress equation (\ref{nongrav1}) in terms of orbital elements.  Doing so
requires us to express $\vec{J}_{\rm NG}$ in terms of positions and velocities.
In general, $D \propto r^{-2}$ \citep{sosfer2009}, as the gas production rate
peaks at the pericentre.  More complex expressions \citep[e.g.][]{koretal2014} are Solar
system-specific, and are dependent on observables.
Also, $\vec{v}_{\rm g} \propto r^{-1/4}$ \citep[equation 28 of][]{whipple1950}.
We note that this velocity dependence is different from the $r^{-1/2}$ dependence
predicted by equation 8 of \citep{koretal2010} because the latter is the terminal
velocity of the gas, rather than the velocity at the area of the ejection.
Assuming that sublimation occurs only on the starlit side, we find
$\vec{v}_{\rm g} / v_{\rm g} = \vec{r} / r$.

These dependencies yield:

\begin{equation}
\ddot{\vec{r}}_{\rm sub} = 
-
\Theta\left(M_{\rm v}\right)
\frac
{ 
D_{0}v_{\rm g0}m_{\rm s}
}
{
M_{\rm c} 
}
\left(
\frac{r_{0}}{r}
\right)^{\frac{9}{4}}
\frac{\vec{r}}{r}
\label{nongrav2}
\end{equation}

\noindent{}where $D = (r_0/r)^{2} D_{0}$ and
$\vec{v}_{\rm g} = v_{\rm g0}(r_0/r)^{\frac{1}{4}} (\vec{r}/r)$.
Although these power-law dependencies are physically-motivated, we realize
that other dependencies may be used based on Solar system observations
\citep[e.g.][]{sanetal2001}.  In order to take into account other possibilities,
we generalise equation (\ref{nongrav2}) in the Appendix.  There we show 
(in Table \ref{TablePower}) 
our qualitative results hold for steeper power-law dependencies.

We cannot assume that the composition and outgassing properties of 
extrasolar minor planets
are similar to Solar system minor planets because their formation channels 
may be different and because WDs and the Sun have different power spectra.
Nevertheless, we can look
towards the Solar system for rough guidance.  In the Solar system,
representative $D_0$ values for sublimation are a few orders of magnitude less than
than the observed outgassing-based values of $D_{0} \sim 10^{28}-10^{31}$ molecules
per second (Table 3 of \citealt*{ahearn1995} and Fig. 3 of \citealt*{sosfer2011}).
The ejection speeds $|\vec{v}_{\rm g}|$ from the surface are unknown, but the
terminal escape speed of the gas is thought to be on the order of cm/s - tens 
of m/s \citep{koretal2010}.  If $r_0$ is taken to be at the pericentre of the orbit, 
then equation (\ref{nongrav2}) effectively 
models the diminution the escaping molecules as the
minor planet flies away from the WD.  This sublimation model is adequate 
for our purposes.  

Real outgassing naturally does not occur in such a well-behaved radially symmetric manner; 
nonzero tangential components may be significant
(e.g. Table 1 of \citealt*{maretal1973}, Table 2 of \citealt*{krolikowska2004}, 
and Table 4 of \citealt*{szuric2006}).  Therefore, the initial conditions for 
exosystems with minor planets and comets are largely unconstrained by 
observationally-motivated considerations.  
Consequently, in Section 4 we consider outgassing from a general analytical perspective
that strengthens our main conclusion without forcing us to resort to specific parameter choices.

Also, for notational ease, we will henceforth drop the $\Theta(M_{\rm v})$ term in
all subsequent equations. Nevertheless, we must determine over what
timescale $M_{\rm v}$ is depleted.  
The volatile budget decreases according to

\begin{eqnarray}
M_{\rm v}
&=& 
M_{\rm v}(0) - 
\int_{0}^{t} 
m_{\rm s} D_0 \left( \frac{r_0}{r} \right)^2 dt
\nonumber
\\
&\approx&
M_{\rm v}(0) 
-
\frac
{m_{\rm s} D_0 r_{0}^2}
{na^2 \sqrt{1-e^2}} 
\left(f-f_0\right)
\label{midm}
\end{eqnarray}

\begin{eqnarray}
&\approx& M_{\rm v}(0)
- 
8.6 \times 10^7 {\rm kg} \times N
\left( \frac{\sqrt{1-0.999^2}}{\sqrt{1-e^2}}\right)
\nonumber
\\
&\ \ \ \times&
\left( \frac{a}{10 {\rm au}}\right)^{-1/2}
\left( \frac{r_0}{10^{-2} {\rm au}}\right)^{2}
\left( \frac{M}{0.6 M_{\odot}}\right)^{-1/2}
\nonumber
\\
&\ \ \ \times&
\left( \frac{m_{\rm s}}{2.99 \times 10^{-26} {\rm kg}}\right)
\left( \frac{D_0}{10^{29}{\rm mol/s}}\right)
\label{massloss}
\end{eqnarray}

\noindent{}where $N$ is the approximate number of orbits, $n$ is the mean motion,
$a$ is the semimajor axis, $e$ is the eccentricity and $f$ is the true anomaly.
The number of orbits over which the entire volatile budget would be depleted
is

\begin{equation}
N_{\rm max} \approx \frac{M_{\rm v}(0)n_0a_{0}^2\sqrt{1-e_{0}^2}}{2\pi m_{\rm s} D_0 r_{0}^2}
.
\label{Nmax}
\end{equation}

\noindent{}Equation (\ref{Nmax}) is useful for determining simulation
durations, and how much of the body's mass has been lost when the body's
orbit begins to undergo major changes.  However, that estimate becomes
progressively worse as the semimajor axis undergoes drastic variations.

Having obtained an observationally-motivated functional form for sublimation 
(equation \ref{nongrav2}), we can 
now derive the equations of motion in orbital elements to understand how 
sublimation changes the orbit.

\section{Unaveraged equations of motion for sublimation}

As a first step, we consider the equations of motion with as few assumptions as possible.
In the subsequent analysis, the following auxiliary set of variables will
be useful:

\begin{eqnarray}
C_1 &\equiv& e \cos{\omega} + \cos{\left(f+\omega\right)}
,
\label{C1}
\\
C_2 &\equiv& e \sin{\omega} + \sin{\left(f+\omega\right)}
,
\label{C2}
\\
C_5 &\equiv& \left(3 + 4e \cos{f} + \cos{2f} \right) \sin{\omega}
        + 2 \left( e + \cos{f} \right) \cos{\omega} \sin{f}
,
\nonumber
\\
&&
\\
C_6 &\equiv& \left(3 + 4e \cos{f} + \cos{2f} \right) \cos{\omega}
        - 2 \left( e + \cos{f} \right) \sin{\omega} \sin{f}
,
\nonumber
\\
&&
\\
C_7 &\equiv& \left(3 + 2e \cos{f} - \cos{2f} \right) \cos{\omega}
        + \sin{\omega} \sin{2f}
,
\\
C_8 &\equiv& \left(3  - \cos{2f} \right) \sin{\omega}
        - 2 \left(e + \cos{f} \right) \cos{\omega} \sin{f} 
,
\\
C_9 &\equiv& \left(3 + 2e \cos{f} - \cos{2f} \right) \sin{\omega}
        - \cos{\omega} \sin{2f}
.
\label{C9}
\end{eqnarray}

\noindent{}This nomenclature was chosen to maintain consistency with
previous studies \citep{vereva2013a,vereva2013b,veretal2013a,veretal2014d,veras2014a}
and highlight the naturally-occurring quantities in many formulations
of the perturbed two-body problem, including those with sublimation.

\subsection{General orbital evolution equations}

In order to obtain the unaveraged equations of motion in orbital elements,
we employ the procedure suggested in \cite{veretal2011} and embellished
in \cite{vereva2013a}, based on perturbative analyses from \cite{efroimsky2005} 
and \cite{gurfil2007}.  We find

\setlength{\belowdisplayskip}{0pt} 
\setlength{\belowdisplayshortskip}{0pt}
\setlength{\abovedisplayskip}{0pt} 
\setlength{\abovedisplayshortskip}{0pt}
\[
\left( \frac{da}{dt} \right)_{\rm sub}
=
\frac
{2 D_{0}v_{\rm g0}m_{\rm s}r_{0}^{9/4} \left(1 + e \cos{f} \right)^{9/4}}
{M_{\rm c} n a^{9/4} \left(1 - e^2 \right)^{11/4}}
\]

\[
\ \ \ \ \ \times \bigg[
C_1 \left(\cos{i}\cos{\Omega} - \cos{i}\sin{\Omega} + \sin{i} \right)
\]

\begin{equation}
\ \ \ \ \ -
C_2 \left(\cos{\Omega} +\sin{\Omega} \right)
\bigg]
,
\label{agen}
\end{equation}

\[
\left( \frac{de}{dt} \right)_{\rm sub}
=
\frac
{D_{0}v_{\rm g0}m_{\rm s}r_{0}^{9/4}\left(1 + e \cos{f} \right)^{5/4} }
{2 M_{\rm c} n a^{13/4} \left(1 - e^2 \right)^{7/4}}
\]

\[
\ \ \ \ \ \times \bigg[
C_6 \left(\cos{i}\cos{\Omega} - \cos{i}\sin{\Omega} + \sin{i} \right)
\]

\begin{equation}
\ \ \ \ \ -
C_5 \left(\cos{\Omega} +\sin{\Omega} \right)
\bigg]
,
\label{egen}
\end{equation}

\[
\left( \frac{di}{dt}  \right)_{\rm sub}
=
\frac
{D_{0}v_{\rm g0}m_{\rm s}r_{0}^{9/4}\left(1 + e \cos{f} \right)^{5/4} \cos{\left(f+\omega\right)} }
{M_{\rm c} n a^{13/4} \left(1 - e^2 \right)^{7/4}}
\]

\begin{equation}
\ \ \ \ \ \times \sin{i} \left(\sin{\Omega} - \cos{\Omega} + \cot{i} \right)
,
\end{equation}

\[
\left( \frac{d\Omega}{dt} \right)_{\rm sub}
=
\frac
{D_{0}v_{\rm g0}m_{\rm s}r_{0}^{9/4}\left(1 + e \cos{f} \right)^{5/4} \sin{\left(f+\omega\right)} }
{M_{\rm c} n a^{13/4} \left(1 - e^2 \right)^{7/4}}
\]

\begin{equation}
\ \ \ \ \ \times \left(\sin{\Omega} - \cos{\Omega} + \cot{i} \right)
,
\end{equation}

\[
\left( \frac{d\omega}{dt} \right)_{\rm sub}
=
-
\frac
{D_{0}v_{\rm g0}m_{\rm s}r_{0}^{9/4}\left(1 + e \cos{f} \right)^{5/4} }
{2 M_{\rm c} n a^{13/4} e \left(1 - e^2 \right)^{7/4}}
\]

\[
\ \ \ \ \ \times
\bigg[
C_8 \left(\cos{i}\cos{\Omega} - \cos{i}\sin{\Omega} \right)  + C_9\sin{i}
\]

\begin{equation}
\ \ \ \ \ +
C_7 \left(\cos{\Omega} +\sin{\Omega} \right)
+
2 e \sin{\left(f+\omega\right)}\cos{i}\cot{i}
\bigg]
,
\label{omgen}
\end{equation}

\setlength{\belowdisplayskip}{5pt} 
\setlength{\belowdisplayshortskip}{5pt}
\setlength{\abovedisplayskip}{5pt} 
\setlength{\abovedisplayshortskip}{5pt}

\[
\left( \frac{dq}{dt} \right)_{\rm sub}
=
\frac
{D_{0}v_{\rm g0}m_{\rm s}r_{0}^{9/4} \left(1 + e \cos{f} \right)^{5/4} }
{M_{\rm c} n a^{9/4} \left(1 - e\right)^{3/4} \left(1 + e\right)^{11/4}}
\sin{\left(\frac{f}{2}\right)}
\]
\setlength{\belowdisplayskip}{0pt} 
\setlength{\belowdisplayshortskip}{0pt}
\setlength{\abovedisplayskip}{0pt} 
\setlength{\abovedisplayshortskip}{0pt}
\[
\ \ \ \ \ \bigg\lbrace
-2\cos{\left(\frac{f}{2}\right)}
\left(2 + e - \cos{f} \right)
\bigg[  
\cos{\omega}  \left(\cos{\Omega} + \sin{\Omega}  \right)
\]

\[
\ \ \ \ \ + \sin{\omega} \left(\sin{i} + \cos{i} \left(\cos{\Omega} - \sin{\Omega} \right) \right)
\bigg]
\]

\[
\ \ \ \ \ +4\sin^3{\left(\frac{f}{2}\right)}
\bigg[  
\sin{\omega}  \left(\cos{\Omega} + \sin{\Omega}  \right)
\]

\begin{equation}
\ \ \ \ \ - \cos{\omega} \left(\sin{i} - \cos{i} \left(\sin{\Omega} - \cos{\Omega} \right) \right)
\bigg]
\bigg\rbrace
,
\label{qgen}
\end{equation}

\begin{equation}
\frac{df}{dt} = 
\frac
{n\left(1 + e \cos{f} \right)^2}
{\left(1 - e^2\right)^{3/2}}
-
\frac{d\omega}{dt}
-
\cos{i}
\frac{d\Omega}{dt}
,
\label{dfdt}
\end{equation}

\noindent{}

\noindent{}where $i$ is the inclination, $\Omega$ is the longitude of ascending node,
$\omega$ is the argument of pericentre and $q$ is the pericentre distance. Note that
the form of equation (\ref{dfdt}), without the subscript ``sub'', is not restricted to 
sublimative forces.

Our main result can be hinted at directly by inspection from 
equations (\ref{agen})-(\ref{dfdt}): for minor planets with orbital 
eccentricities near unity, sublimation-induced changes to
the semimajor axis, eccentricity and pericentre distance are relatively 
strong, minor and negligible,
respectively.
Altering the pericentre distance of an orbit of a highly eccentric minor planet 
around a WD through sublimation alone is difficult.  Other agents in the system 
are needed to perturb the
minor planet into an orbit where it can tidally disrupt.

We can facilitate the coupling of GR with sublimation by considering
the corresponding equations of motion in orbital elements. 
By assuming $M \ll M_{\rm WD}$ to an excellent approximation
we find \citep{veras2014b}

\[
\left( \frac{da}{dt} \right)_{\rm GR}
\approx
\frac
{2G^2 M_{\rm WD}^2 e \sin{f} \left(1 + e \cos{f} \right)^2}
{c^2 n a^{3} \left(1 - e^2 \right)^{7/2}}
\]

\begin{equation}
\ \ \ \ \ \ \ \ \ \ \ \ \ \times
\left[
7+ 3e^2 + 10 e \cos{f}
\right]
,
\label{aGR}
\end{equation}

\[
\left( \frac{de}{dt} \right)_{\rm GR}
\approx
\frac
{G^2 M_{\rm WD}^2 \sin{f} \left(1 + e \cos{f} \right)^2}
{c^2 n a^{4} \left(1 - e^2 \right)^{5/2}}
\]

\begin{equation}
\ \ \ \ \ \ \ \ \ \ \ \ \ \times
\left[
3 + 7e^2 + 10 e \cos{f}
\right]
,
\label{eGR}
\end{equation}

\begin{equation}
\left( \frac{di}{dt} \right)_{\rm GR}
=
\left( \frac{d\Omega}{dt} \right)_{\rm GR}
=
0,
\end{equation}

\noindent{}

\[
\left( \frac{d\omega}{dt} \right)_{\rm GR}
\approx
\frac
{G^2 M_{\rm WD}^2 \left(1 + e \cos{f} \right)^2}
{c^2 n a^{4} e \left(1 - e^2 \right)^{5/2}}
\]

\begin{equation}
\ \ \ \ \ \ \ \ \ \ \ \ \ \times
\left[
\left(e^2-3\right)\cos{f} + 3 e  - 5 e \cos{\left(2f\right)}
\right]
,
\label{omegaGR}
\end{equation}

\[
\left( \frac{dq}{dt} \right)_{\rm GR}
=
\frac
{G^2 M_{\rm WD}^2 \sin{f} \left(1 + e \cos{f} \right)^2}
{c^2 n a^{3} \left(1 - e \right)^{3/2} \left(1 + e \right)^{7/2}}
\]

\begin{equation}
\ \ \ \ \ \ \ \ \ \ \ \ \ \times
\left[
-3 + 8 e + e^2 - 10 e \cos{f}
\right]
.
\label{qGR}
\end{equation}

\subsection{Planar orbital evolution equations}

Equations (\ref{agen})-(\ref{qgen}) demonstrate how sublimation affects
the entire set of orbital elements, including the inclination and longitude
of ascending node.  However, because WDs are extremely 
spherical compared to planets, 
the orbital architecture is rotationally symmetric.  Hence, 
future WD studies might need to consider just the planar case.
In order to obtain the equations for the planar case, 
we take the limit of zero inclination,
and define the longitude of pericentre $\varpi = \Omega + \omega$. 

For the planar equations of motion, the $C$ variables which contain a subscript of 
``P'' (below) are equivalent to those variables defined in equations (\ref{C1})-(\ref{C9})
except that $\omega$ is replaced with $\varpi$.  With these definitions, we can
compactly express the coplanar equations of motion as

\begin{equation}
\left( \frac{da}{dt} \right)_{\rm sub}^{\rm P}
=
\frac
{2 D_{0}v_{\rm g0}m_{\rm s}r_{0}^{9/4} \left(1 + e \cos{f} \right)^{9/4} }
{M_{\rm c} n a^{9/4} \left(1 - e^2\right)^{11/4}   }
\left(
C_{\rm 1P} - C_{\rm 2P} 
\right)
,
\label{dadtP}
\end{equation}

\begin{equation}
\left( \frac{de}{dt} \right)_{\rm sub}^{\rm P}
=
\frac
{D_{0}v_{\rm g0}m_{\rm s}r_{0}^{9/4} \left(1 + e \cos{f} \right)^{5/4} }
{2 M_{\rm c} n a^{13/4} \left(1 - e^2\right)^{7/4}  }
\left(
C_{\rm 6P} - C_{\rm 5P} 
\right)
,
\label{dedtP}
\end{equation}

\begin{equation}
\left( \frac{d\varpi}{dt} \right)_{\rm sub}^{\rm P}
=
-
\frac
{D_{0}v_{\rm g0}m_{\rm s}r_{0}^{9/4} \left(1 + e \cos{f} \right)^{5/4} }
{2 M_{\rm c} n e a^{13/4} \left(1 - e^2\right)^{7/4}   }
\left(
C_{\rm 7P} + C_{\rm 9P} 
\right)
,
\label{dcpidtP}
\end{equation}

\begin{equation}
\left( \frac{dq}{dt} \right)_{\rm sub}^{\rm P}
=
-\frac{D_{0}v_{\rm g0}m_{\rm s}r_{0}^{9/4} \left(1 + e \cos{f} \right)^{5/4}  }
{2 M_{\rm c} n a^{9/4} \left(1 - e\right)^{3/4}  \left(1 + e\right)^{11/4}}
\nonumber
\end{equation}

\[
\ \ \ \ \ \times
\bigg[
\cos{\left(2f+\varpi\right)}
+
\cos{\varpi}
\left(
3 - 4 \cos{f} + 2 e \sin{f}
\right)
\]

\begin{equation}
\ \ \ \ \ 
-
\left(
3
-
2\left(2+e\right)
\sin{f}
\right)
\sin{\varpi}
+
4 \sin{\left(f+\varpi\right)}
-
\sin{\left(2f+\varpi \right)}
\bigg]
.
\label{dqdtP}
\end{equation}

\noindent{}

Equations (\ref{dadtP})-(\ref{dqdtP}) show that
even in the planar case, the orbital equations are still nontrivial 
functions of the true anomaly of the
asteroid.  Nevertheless, we can glean some physical insight
from the formulae.  The eccentricity terms in the denominator
illustrate that $(da/dt)_{\rm sub}$ will evolve more quickly
than $(de/dt)_{\rm sub}$,  $(d\varpi/dt)_{\rm sub}$, and 
$(dq/dt)_{\rm sub}$.

For GR, because equations (\ref{aGR})-(\ref{eGR}) and (\ref{omegaGR})-(\ref{qGR})
are all independent of $i$, $\Omega$ and $\omega$, we have

\begin{equation}
\left( \frac{da}{dt} \right)_{\rm GR}^{\rm P}
=
\left( \frac{da}{dt} \right)_{\rm GR}
,
\end{equation}

\begin{equation}
\left( \frac{de}{dt} \right)_{\rm GR}^{\rm P}
=
\left( \frac{de}{dt} \right)_{\rm GR}
,
\end{equation}

\begin{equation}
\left( \frac{dq}{dt} \right)_{\rm GR}^{\rm P}
=
\left( \frac{dq}{dt} \right)_{\rm GR}
,
\end{equation}

\begin{equation}
\left( \frac{d\varpi}{dt} \right)_{\rm GR}^{\rm P}
=
\left( \frac{d\omega}{dt} \right)_{\rm GR}
.
\end{equation}

\subsection{Numerical integrations}

In this section we provide some specific examples of minor planet
orbital evolution due to sublimation and GR.

\subsubsection{Setup}

First, we must establish some fiducial parameters.  A crucial
parameter is $D_0$, which varies by several orders of magnitude
in the Solar system.  This variation should be even higher in
WD systems because WD luminosities and spectral properties
vary by orders of magnitude
depending on the WD cooling age; WDs may be either more luminous
or (much) less luminous than the Sun. 
Although the latter case
holds in the majority of cases, $D_0$ cannot necessarily also be
assumed to be lower because minor planets can reach much smaller
values of $q$ in WD systems.  In fact, without being engulfed, a minor planet
may attain $q = 6 \times 10^{-5}$ au.  However, this value is well
within the WD disruption radius, where the minor planet would likely
be shorn apart.  Instead, $q = 0.02$ au would nearly guarantee that
the minor planet survives; actually survival is likely to
be ensured over hundreds of orbits for values of $q$ which
are one or two orders of magnitude lower \citep{veretal2014a}.
Complicating these considerations for determining a fiducial
$D_0$ is that sublimation is a function of temperature and pressure,
and significantly localized physical properties such as shadowing,
thermal inhomogeneities, and spin.

 \begin{figure*}
  {\huge \bf Without GR \ \ \ \ \ \ \ \ \ \ \ \ \ \ \ 
                    \ \ \ \,  With GR}
   \vskip 0.2cm
   \centerline{
   \psfig{figure=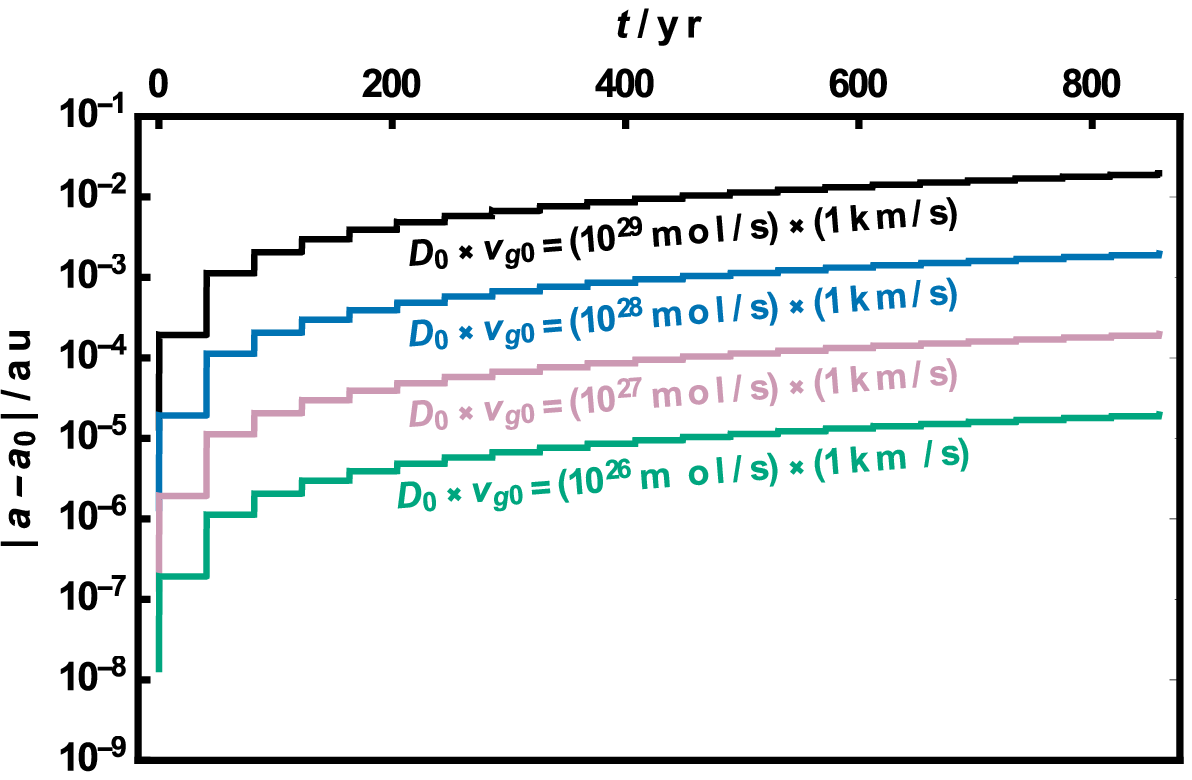,width=8cm}
   \psfig{figure=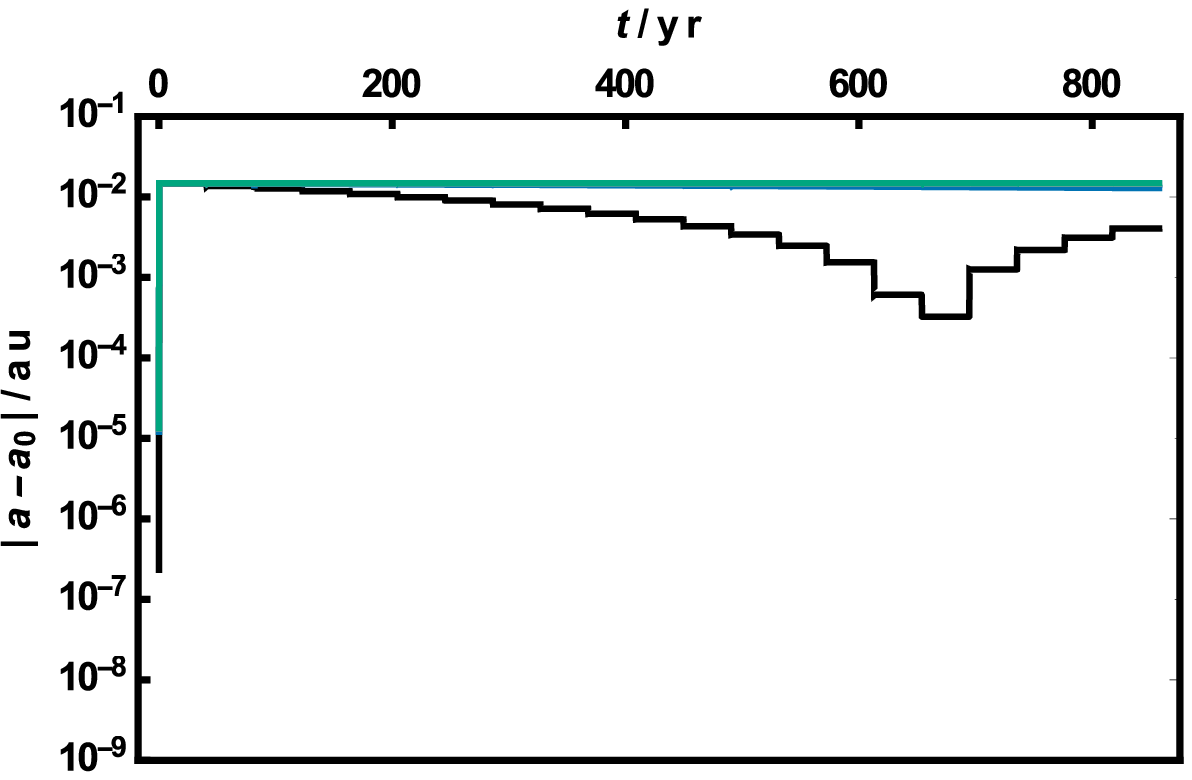,width=8cm}
   }
   \vskip -0.4cm 
   \centerline{
   \psfig{figure=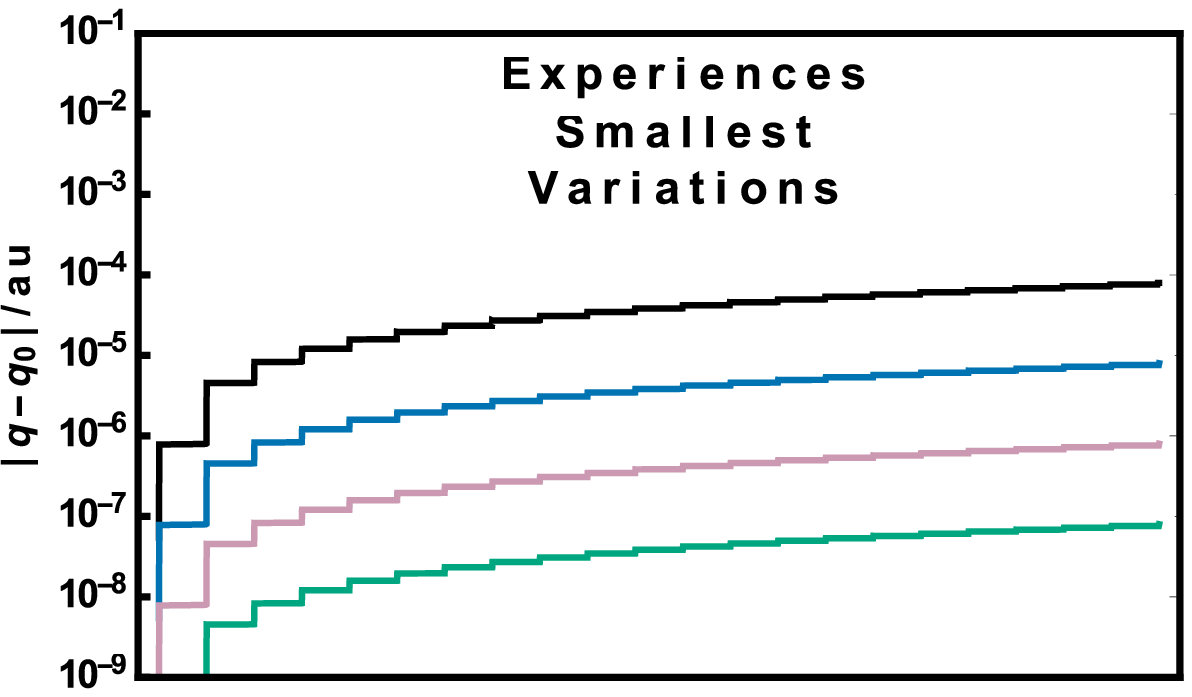,width=8cm}
   \psfig{figure=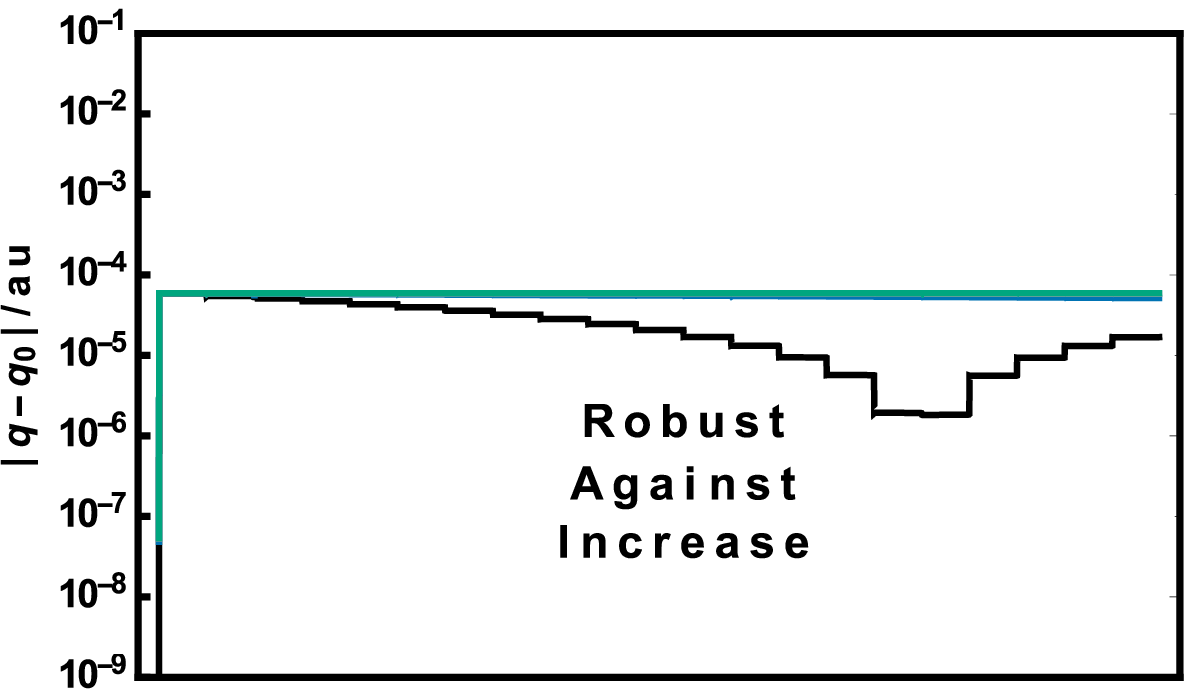,width=8cm}
   }
   \vskip -0.4cm 
   \centerline{
   \psfig{figure=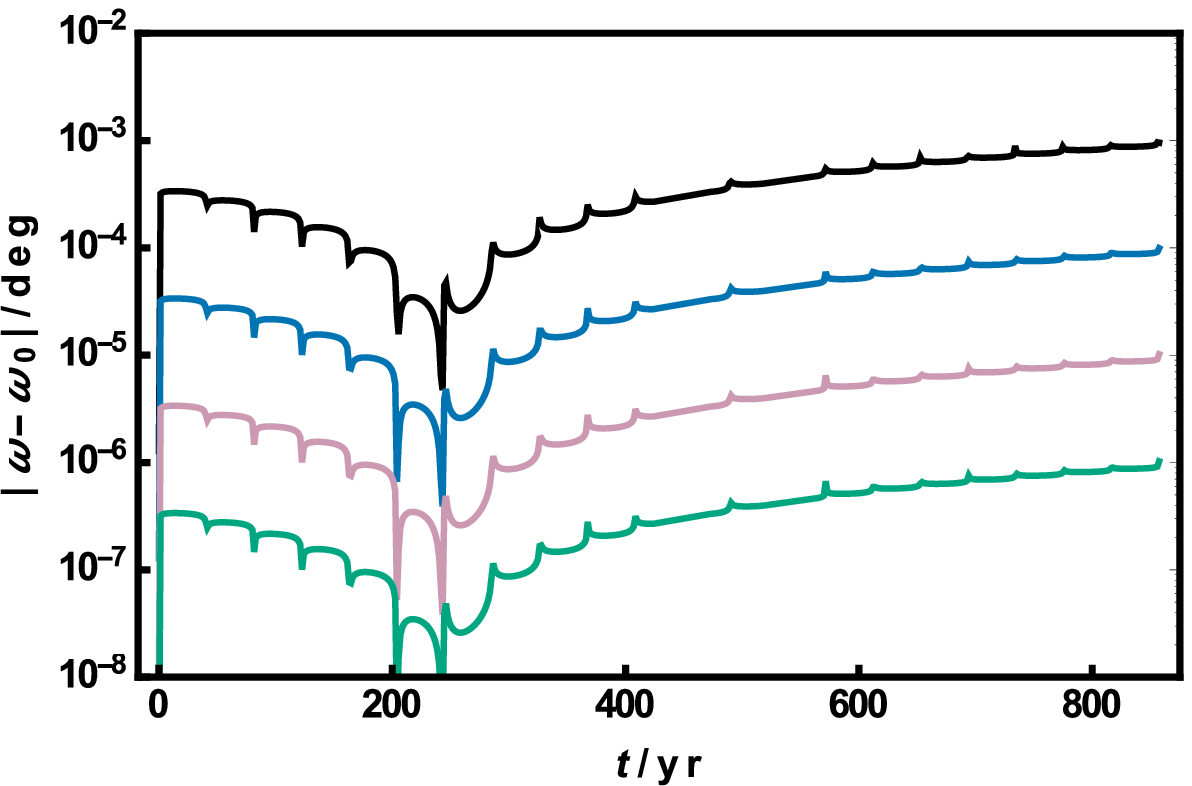,width=8cm}
   \psfig{figure=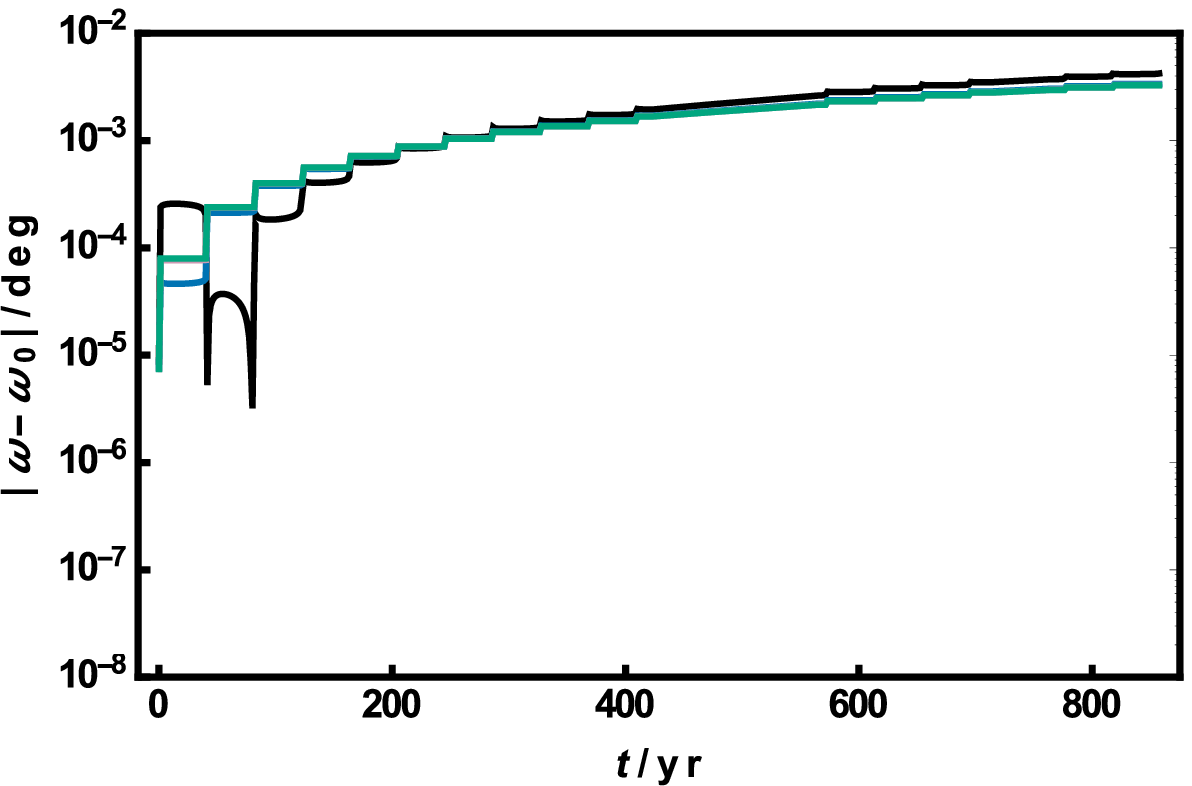,width=8cm}
   }
 \caption{
 The orbital changes over about 25 orbits experienced by a 
 $10^{13}$ kg minor planet orbiting
 a $0.6 M_{\odot}$ star (like a WD) and sublimating water ice at a given
 rate $D_0$ and with a given speed $v_{\rm g0}$.  The product of $D_0$ and 
 $v_{\rm g0}$ scales linearly with the change in each orbital element.
 The left panel does not include GR, and the right panel does.  
 We set the initial minor planet orbit at
 $a_0 = 10$ au, $q_0 = 0.02$ au, $e_0 = 0.998$, $i_0 = 1.0^{\circ}$,
 $\Omega_0 = 180^{\circ}$, $\omega_0 = 1.0^{\circ}$ and $f_0 = 0.0^{\circ}$.
 The plot demonstrates the importance of including GR when the pericentre
 is close to the star, and hints that $q$ cannot drift into the star's
 disruption radius by the combined forces from sublimation and GR alone.
 }
 \label{set1}
 \end{figure*}

 \begin{figure*}
 {\huge \bf With GR }
  \vskip 0.2cm
  \centerline{
   \psfig{figure=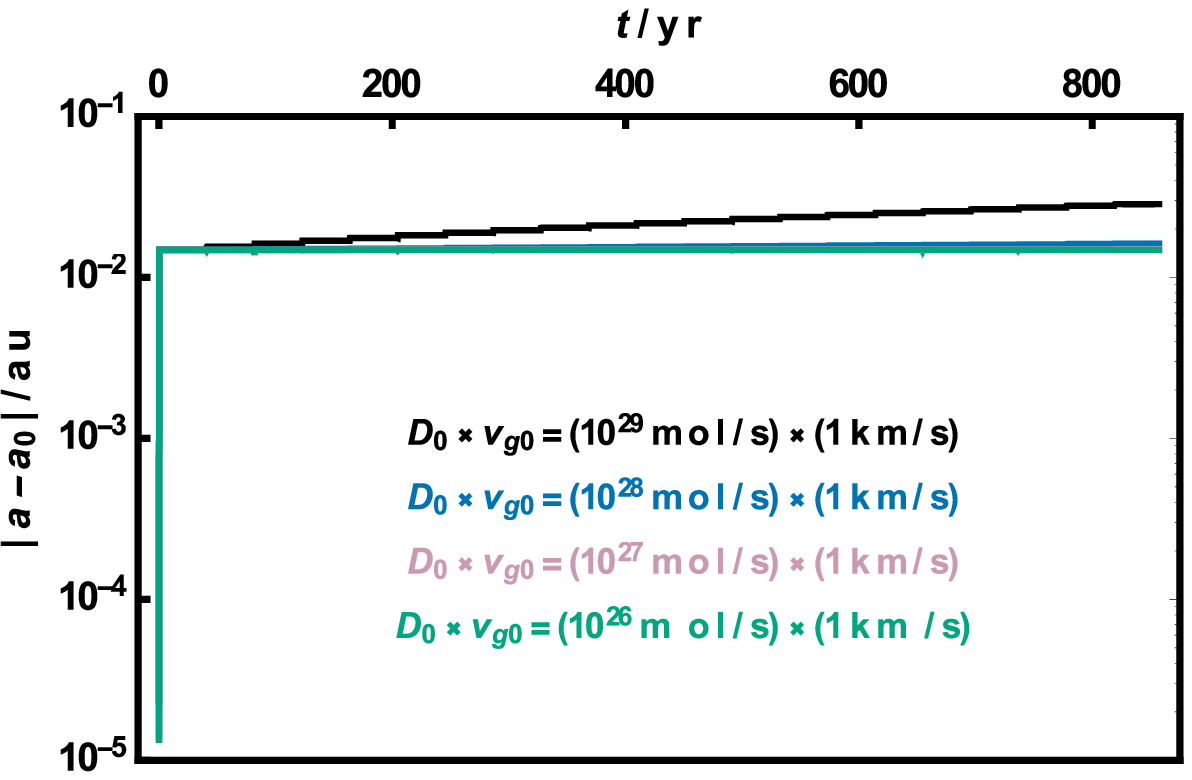,width=8cm}
   \psfig{figure=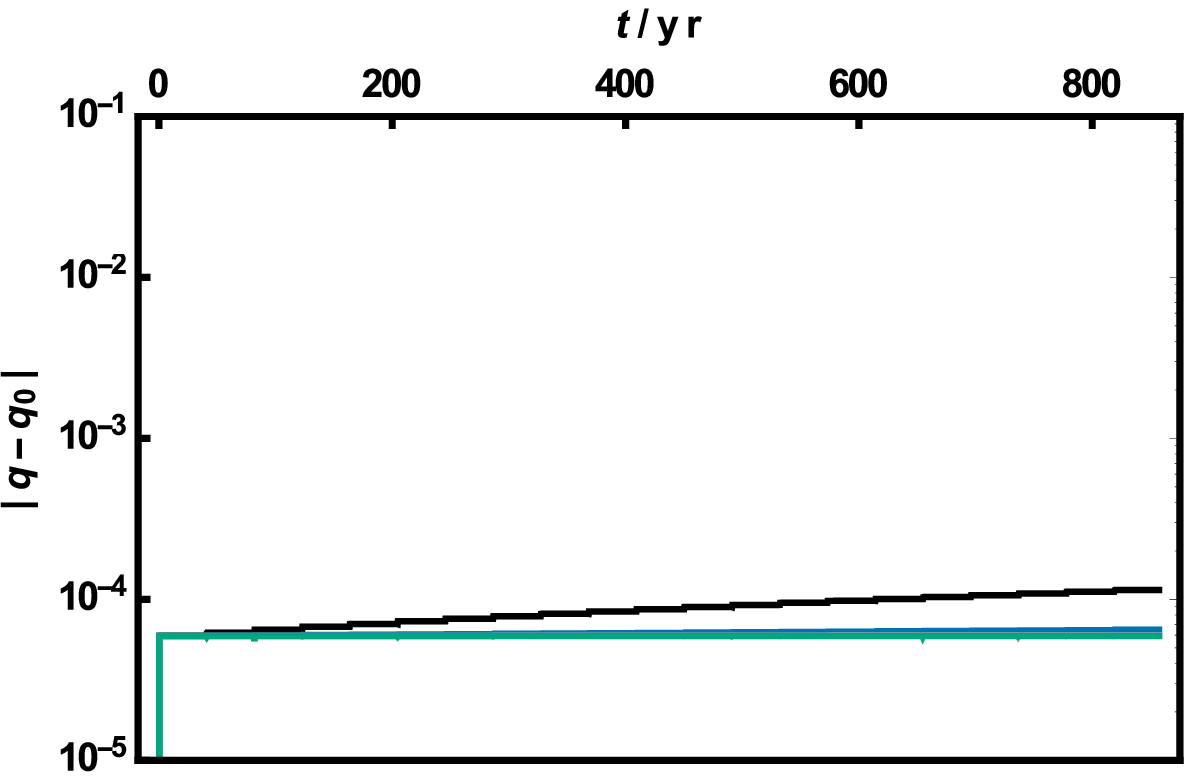,width=8cm}
   }
   \vskip -0.4cm 
   \centerline{
   \psfig{figure=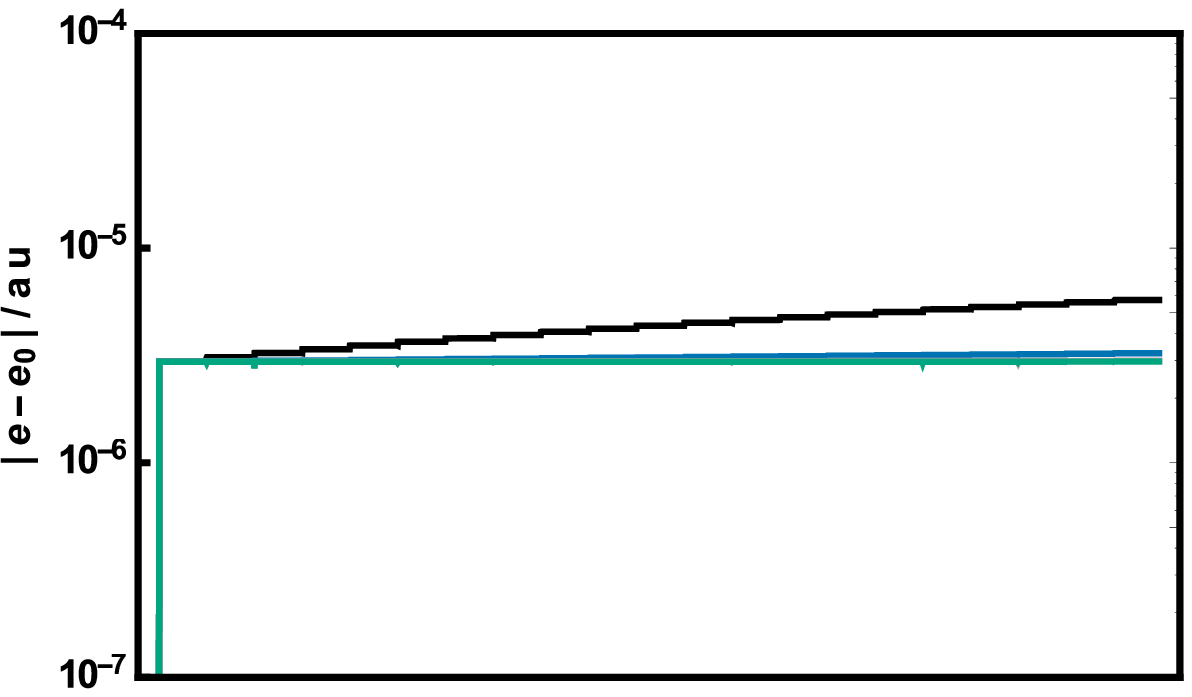,width=8cm}
   \psfig{figure=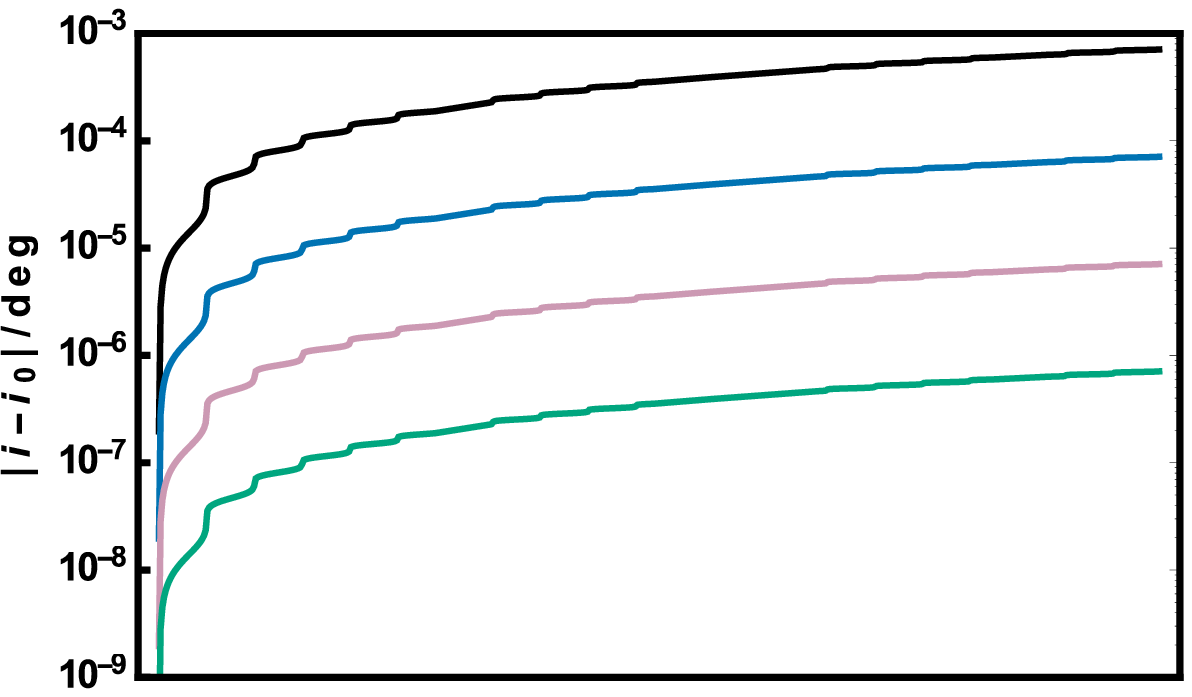,width=8cm}
   }
   \vskip -0.4cm 
   \centerline{
   \psfig{figure=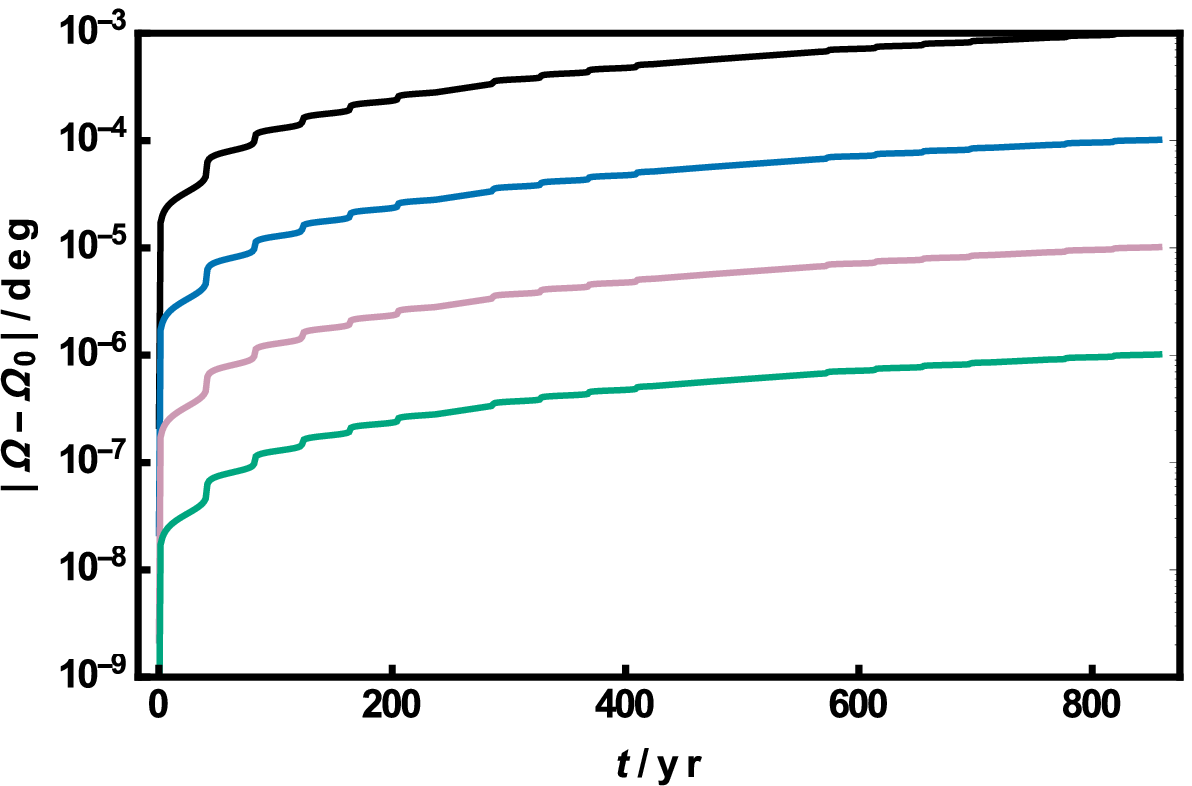,width=8cm}
   \psfig{figure=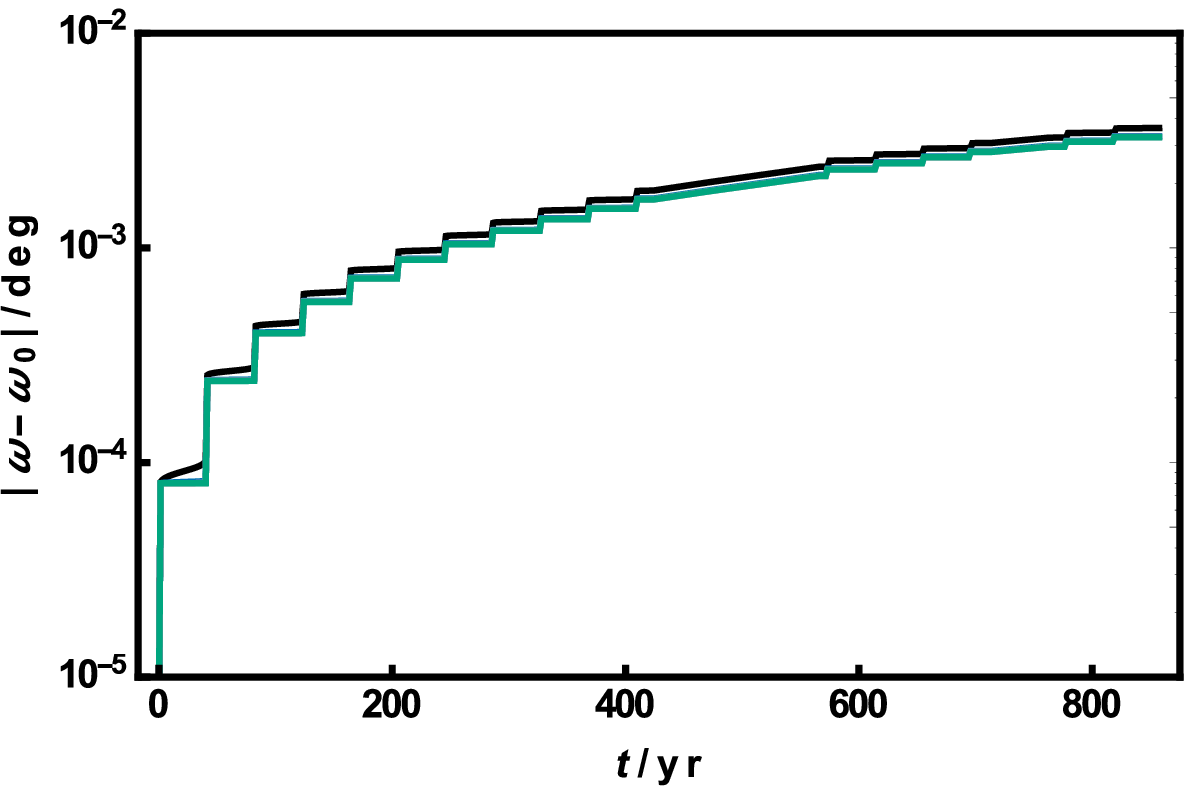,width=8cm}
  }
 \caption{
 Like Figure \ref{set1}, except with  $i_0 = 45^{\circ}$,
 $\Omega_0 = 200^{\circ}$ and $\omega_0 = 135^{\circ}$. 
 Both sublimation and GR are included in the simulation.
 Illustrated here are the time evolution of all of the orbital elements
 except $f$.  Despite the minor planets having highly inclined
 orbits, $q$ remains robust to major orbital changes.
  }
 \label{set2}
 \end{figure*}

Consequently, we defer linking $D_0$ with detailed internal 
minor planet models and WD cooling age to other studies.  
Instead, here we identify the value of $D_0$ in which sublimative-induced 
orbital perturbations becomes significant for a minor planet
which does not initially quite reach the WD disruption radius (for example,
$q_0 = 0.02$ au).  Like $D_0$, the value of $v_{\rm g0}$ is 
intrinsically linked to local properties of the minor planet
and is unknown outside of the Solar system.  Fortunately, we
can partially circumvent our ignorance by recognizing that
all of the sublimation-based orbital evolution equations 
are directly proportional to the product $D_0 v_{\rm g0}$.
This property allows us to reduce the number of degrees of
freedom in any phase space exploration.  
For our explorations,
we choose the four products corresponding to the range
$D_0 v_{\rm g0} = \left(10^{26-29} {\rm mol}/{\rm s} \right) \times \left(1 {\rm km}/{\rm s}\right)$.
These speeds are easily achieved; in the Solar system, assuming a Maxwell-Boltzmann 
distribution, one obtains $\langle v_{\rm g0} \rangle = 675 \pm 275$ m/s (with extreme speeds of several km/s) 
for temperatures of 360 K \citep{groetal2013}. 

Our other adopted values include 
$M_{\rm WD} = 0.6M_{\odot}$ (representing a typical WD; see 
\citealt*{lieetal2005,camenzind2007,faletal2010,treetal2013}),
and $M_{\rm c} = 10^{13}$ kg (roughly corresponding
to an active asteroid with a radius of 1 km and density of 2 g/cm$^3$).
We place our minor planets on $a_0 = 10$ au orbits, because most
km-sized minor planets within about 7 au will have been spun up
to fission from giant branch radiation \citep{veretal2014b}.
We do adopt $q = 0.02$ au, meaning that $e_0 = 0.998$, a typical
eccentricity for a minor planet just missing the WD disruption radius.
We sample minor planets on both slightly inclined ($\sim 1^{\circ}$;
Figure \ref{set1}) and highly inclined ($\sim 45^{\circ}$;
Figure \ref{set2}) orbits with respect to the same fixed but
arbitrary reference plane. 
Our simulations begin at the pericentre
and we propagate the systems for approximately 25 orbits.  We integrate
equations (\ref{agen})-(\ref{omgen}) and (\ref{dfdt})-(\ref{omegaGR})
all together.

\subsubsection{Results}

We present our results in Figs. \ref{set1}-\ref{set2}.  The sawtooth-like
curves in both figures indicate that sublimation causes the greatest
orbital changes at the pericentre, so that each step corresponds to one
orbit.

Figure \ref{set1} showcases a side-by-side comparison when GR is 
not included (left panels) 
and is included (right panels) in the simulations.  The difference is 
striking, particularly for the semimajor axis and pericentre evolution; 
for the blue, pink and green curves, at least, GR dominates over sublimation.  
Alternatively, the black curve for the evolution
of $\omega$ indicates a regime 
where sublimation may dominate GR (after a few hundred yr).  
The figure makes clear that for reasonable parameters that one would 
expect in a WD system with planetary remnants, GR can play a crucial 
role and should be included in sublimation models.

Another comparison to consider in that figure is between the top and 
middle plots, whose $y$-axes are equivalent and in au.  The change in 
$q-q_0$ is typically orders of magnitude less than the change in 
$a-a_0$, both with and without GR.  GR dominates the evolution of 
$q$ in most cases, flattening out the curves.  In no case does
$q$ vary by more than $10^{-4}$ au.

In Figure \ref{set2}, we include GR and show the time evolution
for $e$, $i$, and $\Omega$ and as well as for $a$, $q$ and $\omega$, 
but now for minor planets 
with different initial values of $i$, $\Omega$ and $\omega$.  The 
highly inclined retrograde orbits sampled here yield different 
evolutions from Figure 1, but reinforce the main result: 
the pericentre change does not exceed $2 \times 10^{-4}$ au for any curve,
whereas the semimajor axis may vary by as much as 0.05 au.
Note that the two uppermost plots are drawn on the same scale.

For both sets of simulations, we also computed the total amount of mass actually
lost by the minor planet
from equation (\ref{midm}) throughout the evolution.  This value is 
about $9.7 \times 10^8$ kg, four orders of magnitude less than than
the core mass of the minor planet.

\section{Transverse and Nonplanar Outgassing}

So far our model for sublimation assumes that the nongravitational 
acceleration acts opposite to the radial direction.
But what if the minor planet experiences a significant outgassing episode
in another direction?  In order to answer this question, we first 
consider a commonly-used formulation for outgassing in the Solar system.

\subsection{Sublimation in the Solar system}

This prescription effectively models a perturbed two-body 
problem, where the perturbative force is the product of a 
semiempirical functional form for water ice 
\citep{delmil1971} and a linear combination of the radial, transverse and normal
components of the motion \citep{maretal1973}.
Equation 1 of \cite{sosfer2011} and 
equation 1 of \cite{szutowicz2000} demonstrate 
that the Marsden formulation of outgassing for Solar system comets 
may be expressed as 

\[
\ddot{\vec{r}}_{\rm out} = -\frac{G\left(M_{\odot} + M_{\rm c}\right)}{r^3}
\]

\begin{equation}
\ \ \ \ 
+ \mathbb{N}(r)
\left[
A_1 
\underbrace{\frac{\vec{r}}{r}}_{\begin{subarray}{c}{\rm Radial} \\ \equiv \hat{R} \end{subarray}}
+
A_2 
\underbrace{
\frac
{r \vec{v} - \vec{r} \left(\frac{\vec{r} \boldsymbol{\cdot} \vec{v}}{r} \right) }
{\left| \vec{r} \boldsymbol{\times} \vec{v} \right|}
}_{\begin{subarray}{c}{\rm Transversal} \\ \equiv \hat{T} \end{subarray}}
+
A_3 
\underbrace{
\frac
{\vec{r} \boldsymbol{\times} \vec{v} }
{\left| \vec{r} \boldsymbol{\times} \vec{v} \right|}
}
_{\begin{subarray}{c}{\rm Normal} \\ \equiv \hat{N} \end{subarray}}
\right]
,
\label{rtn}
\end{equation}

\begin{equation}
\mathbb{N}(r)
\equiv
\alpha \left(\frac{r}{r_0}\right)^{-\eta}
\left[1 + \left(\frac{r}{r_0}\right)^{\xi} \right]^{-\zeta}
\end{equation}

\noindent{}

\noindent{}such that the constants $\eta$, $\xi$ and $\zeta$ are empirically
derived, with typical fiducial values of 2.15, 5.093 and 4.6142, respectively. Further,
the values of $A_1$, $A_2$ and $A_3$ are not constant with time; for real comets,
they can vary by an order of magnitude or more (see Table 2 of \citealt*{froric1986}
and Table 2 of \citealt*{cheyeo2005}). Even 
this formulation, however, cannot explain many observational phenomena.
For example, sometimes outgassing is not symmetric about the pericentre.
Equation 5 of \citet[][]{krolikowska2004} takes into account such cases 
with related functional forms.

Note that we recover the formulation used in Section 3 by setting 
$A_2 = A_3 = 0$, $\eta = 9/4$, $\zeta = 0$ and $\alpha = \Theta\left(M_{\rm v}\right)D_0v_{\rm g0}m_{\rm s}/M_{\rm c}$.
We find the equation for the semimajor axis variation from the Marsden formulation to be

\begin{equation}
\left( \frac{da}{dt} \right)_{\rm out}
=
\mathbb{N}(r)
\times
\frac{2 \left[A_1 e \sin{f} + A_2 \left(1 + e \cos{f}\right) \right]}
{n \left(1 - e^2 \right)^{1/2}}
.
\label{subdadt}
\end{equation}

\noindent{}

\noindent{}This form is equivalent to equation (24) of \cite{burns1976} 
(and equation (2.145) of \citealt*{murder1999}), but without the factor
of $\mathbb{N}(r)$, even though we made no assumption
about the magnitude of the nongravitational force (the force
in those other studies is assumed to be small).  
The factor of $\mathbb{N}(r)$ does not alter the classical functional 
form obtained due to the radial, transverse and normal
components. The evolution of the other orbital parameters are as
follows and can also be re-expressed in the forms present in
\cite{burns1976}.

\[
\left( \frac{de}{dt} \right)_{\rm out}
=
\mathbb{N}(r)
\times
\frac{\sqrt{1-e^2}}{2na\left(1 + e\cos{f}\right)}
\]

\begin{equation}
\times \bigg[
2 A_1 \sin{f} \left(1 + e \cos{f}\right)
+
A_2 \left(4 \cos{f} + e \left(3 + \cos{\left(2f\right)}  \right)  \right)
\bigg]
,
\label{subdedt}
\end{equation}

\begin{equation}
\left( \frac{di}{dt} \right)_{\rm out}
=
\mathbb{N}(r)
\times
A_3 
\frac
{\sqrt{1-e^2} \cos{\left(f + \omega\right)}}
{a n \left(1 + e \cos{f} \right)}
,
\end{equation}

\begin{equation}
\left( \frac{d\Omega}{dt} \right)_{\rm out}
=
\mathbb{N}(r)
\times
A_3 
\frac
{\sqrt{1-e^2} \csc{i} \sin{\left(f + \omega\right)}}
{a n \left(1 + e \cos{f} \right)}
,
\end{equation}

\[
\left( \frac{d\omega}{dt} \right)_{\rm out}
=
-
\mathbb{N}(r)
\times
\frac{\sqrt{1-e^2}}{na}
\]

\begin{equation}
\times \bigg[
A_1 \frac{\cos{f}}{e}
-
A_2 
\frac{\sin{f} \left(2 + e \cos{f} \right)}
{e \left(1 + e \cos{f} \right)} 
+
A_3 \frac{\cot{i} \sin{\left(f + \omega\right)}}
{1 + e \cos{f}}
\bigg]
.
\end{equation}

\noindent{}

From equations (\ref{subdadt})-(\ref{subdedt}), we derive the change in
pericentre $q$ and apocentre $Q$ as 

\[
\left( \frac{dq}{dt} \right)_{\rm out}
=
\mathbb{N}(r)
\times
\frac{\left(1-e\right)^2}{2n\sqrt{1-e^2}\left(1 + e\cos{f}\right)}
\]

\begin{equation}
\times \bigg[
-2 A_1 \sin{f} \left(1 + e \cos{f}\right)
+
A_2 \left(4 + e - 4\cos{f} - e \cos{\left(2f\right)} \right)
\bigg]
,
\label{subdqdt}
\end{equation}

\[
\left( \frac{dQ}{dt} \right)_{\rm out}
=
\mathbb{N}(r)
\times
\frac{\left(1+e\right)^{3/2}}{2n\sqrt{1-e}\left(1 + e\cos{f}\right)}
\]

\begin{equation}
\times \bigg[
2 A_1 \sin{f} \left(1 + e \cos{f}\right)
+
A_2 \left(4 - e + 4\cos{f} + e \cos{\left(2f\right)} \right)
\bigg]
.
\label{subdqdt2}
\end{equation}

\noindent{}

We see that $da/dt, de/dt, dq/dt$ and $dQ/dt$ are unaffected
by the choice of $A_3$. If $A_3=0$ then
the inclination and the longitude of the ascending node stay constant.
If the motion is fully co-planar, i.e. $A_3=0$ and $i=0$, then the ascending node 
is no longer well-defined and the evolution of the longitude of pericentre becomes

\[
\left( \frac{d\varpi}{dt} \right)_{\rm out}^{\rm P}
=
\mathbb{N}(r)
\times
\frac{\sqrt{1-e^2}}{aen \left(1 + e \cos{f} \right)}
\]

\begin{equation}
\times \bigg[
-A_1 \cos{f} \left(1 + e \cos{f}\right)
+A_2 \sin{f} \left(2 + e \cos{f}\right)
\bigg]
.
\label{subdvpidt}
\end{equation}

\noindent{}

\noindent{}The formulae in this section allow one to model
the evolution of a particular exosystem minor body for an assumed set
of $\left\lbrace A_1,A_2,A_3 \right\rbrace$.  Currently, no
such sets have been observationally measured outside of the Solar
system.

\subsection{Invariance of the pericentre}

Regardless, we can reveal important properties of the motion
by analyzing the equations alone.
We know that orbital changes predominately occur during pericentre
passages; consequently, we can expand the equations about $f = 0^{\circ}$.
Further, in WD systems, any minor planets or comets which closely 
approach the WD must have
extremely eccentric orbits ($e \ge 0.998$).  Therefore, we can also
Taylor expand the equations about $\epsilon = 1-e$, where $\epsilon \ll 1$.
The following result shows the explicit dependencies on $\left(1 - e\right)$,
$A_1$, $A_2$ and $A_3$.  We facilitate the expressions by first defining


\begin{equation}
\mathcal{N}
\equiv
\alpha \left(\frac{a\left(1-e\right)}{r_0}\right)^{-\eta}
\left[1 + \left(\frac{a\left(1-e\right)}{r_0}\right)^{\xi} \right]^{-\zeta}
.
\end{equation}

\noindent{}


\noindent{}We obtain

\[
\left( \frac{da}{dt} \right)_{\rm out} \bigg|_{\begin{subarray}{c}{f \ {\rm about} \ 0^{\circ}} \\\epsilon \ {\rm about} \ 0 \end{subarray}}
=
\]

\[
\ \ \ \ \ \ \ \
\mathcal{N}
\left\lbrace
-\frac{2\sqrt{2}}{n} \left(1 - e\right)^{-\frac{1}{2}}A_2 + \mathcal{O}\left[ \left(1-e\right)^{\frac{1}{2}} \right]
\right\rbrace
\]

\begin{equation}
\ \ \ \ \ \
+ \mathcal{N}
\left\lbrace
-\frac{\sqrt{2}}{n} \left(1 - e\right)^{-\frac{1}{2}}A_1 + \mathcal{O}\left[ \left(1-e\right)^{\frac{1}{2}} \right] 
\right\rbrace
f
+ \mathcal{O}\left[f^2\right]
.
\end{equation}

\noindent{}

Hence, the semimajor axis is most strongly influenced by a nonzero $A_2$ term, at the 
$\left(1-e\right)^{-1/2}$ order.  The next strongest influence comes from a 
nonzero $A_1$ term at the $f\left(1-e\right)^{-1/2}$ order.  The eccentricity dependencies
are similar, except weaker by a factor of $\left(1-e\right)$:


\noindent{}

\[
\left( \frac{de}{dt} \right)_{\rm out} \bigg|_{\begin{subarray}{c}{f \ {\rm about} \ 0^{\circ}} \\\epsilon \ {\rm about} \ 0 \end{subarray}}
=
\]

\[
\ \ \ \ \ \ \ \
\mathcal{N}
\left\lbrace
-\frac{2\sqrt{2}}{an} \left(1 - e\right)^{\frac{1}{2}}A_2 + \mathcal{O}\left[ \left(1-e\right)^{\frac{3}{2}} \right]
\right\rbrace
\]

\begin{equation}
\ \ \ \ \ \ \
+ \mathcal{N}
\left\lbrace
-\frac{\sqrt{2}}{an} \left(1 - e\right)^{\frac{1}{2}}A_1 + \mathcal{O}\left[ \left(1-e\right)^{\frac{3}{2}} \right] 
\right\rbrace
f 
+ \mathcal{O}\left[f^2\right]
.
\end{equation}

\noindent{}

The inclination and longitude of ascending node terms are independent of $A_1$ and $A_2$.  Instead,
these orbital parameters vary due to a nonzero $A_3$, at the $\left(1-e\right)^{1/2}$ leading order as follows


\noindent{}

\[
\left( \frac{di}{dt} \right)_{\rm out} \bigg|_{\begin{subarray}{c}{f \ {\rm about} \ 0^{\circ}} \\\epsilon \ {\rm about} \ 0 \end{subarray}}
=
\]

\[
\ \ \ \ \ \ \ \
\mathcal{N}
\left\lbrace
-\frac{\cos{\omega}}{\sqrt{2}an} \left(1 - e\right)^{\frac{1}{2}}A_3 + \mathcal{O}\left[ \left(1-e\right)^{\frac{3}{2}} \right]
\right\rbrace
\]

\begin{equation}
\ \ \ \ \ \ \
+ \mathcal{N}
\left\lbrace
\frac{\sin{\omega}}{\sqrt{2}an} \left(1 - e\right)^{\frac{1}{2}}A_3 + \mathcal{O}\left[ \left(1-e\right)^{\frac{3}{2}} \right] 
\right\rbrace
f
+ \mathcal{O}\left[f^2\right]
,
\end{equation}


\noindent{}

\[
\left( \frac{d\Omega}{dt} \right)_{\rm out} \bigg|_{\begin{subarray}{c}{f \ {\rm about} \ 0^{\circ}} \\\epsilon \ {\rm about} \ 0 \end{subarray}}
=
\]

\[
\ \ \ \ \ \ \ \
\mathcal{N}
\left\lbrace
-\frac{\sin{\omega}}{\sqrt{2}an\sin{i}} \left(1 - e\right)^{\frac{1}{2}}A_3 + \mathcal{O}\left[ \left(1-e\right)^{\frac{3}{2}} \right]
\right\rbrace
\]

\begin{equation}
\ \ \ \ \ \ \ \ \
+ 
\mathcal{N}
\left\lbrace
-\frac{\cos{\omega}}{\sqrt{2}an\sin{i}} \left(1 - e\right)^{\frac{1}{2}}A_3 + \mathcal{O}\left[ \left(1-e\right)^{\frac{3}{2}} \right] 
\right\rbrace
f
+ \mathcal{O}\left[f^2\right]
.
\end{equation}

\noindent{}

Unlike the evolution of $a$, $e$, $i$, and $\Omega$, the evolution of $\omega$ is instead dominated by
the combination of an $A_1$ term and a $A_3$ term, as follows


\noindent{}

\[
\left( \frac{d\omega}{dt} \right)_{\rm out} \bigg|_{\begin{subarray}{c}{f \ {\rm about} \ 0^{\circ}} \\\epsilon \ {\rm about} \ 0 \end{subarray}}
=
\]

\[
\ \ \ \ \ \ \ \
\mathcal{N}
\bigg\lbrace
\frac{2}{\sqrt{2}an} \left(1 - e\right)^{\frac{1}{2}}A_1 + \frac{\cot{i}\sin{\omega}}{\sqrt{2}an} \left(1 - e\right)^{\frac{1}{2}}A_3 
\]

\[
\ \ \ \ \ \ \ \ \ \ \
+ 
\mathcal{O}\left[ \left(1-e\right)^{\frac{3}{2}} \right]
\bigg\rbrace
\]

\[
\ \ \ \ \ \ \ \
+ 
\mathcal{N}
\bigg\lbrace
-\frac{3}{\sqrt{2}an} \left(1 - e\right)^{\frac{1}{2}}A_2 + \frac{\cot{i}\cos{\omega}}{\sqrt{2}an} \left(1 - e\right)^{\frac{1}{2}}A_3 
\]

\begin{equation}
\ \ \ \ \ \ \ \ \ \ \
+ \mathcal{O}\left[ \left(1-e\right)^{\frac{3}{2}} \right]
\bigg\rbrace
f
+ \mathcal{O}\left[f^2\right]
.
\end{equation}

\noindent{}

\noindent{}The coplanar version is


\noindent{}

\[
\left( \frac{d\varpi}{dt} \right)_{\rm out} \bigg|_{\begin{subarray}{c}{f \ {\rm about} \ 0^{\circ}} \\\epsilon \ {\rm about} \ 0 \end{subarray}}
=
\]

\[
\ \ \ \ \ \ \ \
\mathcal{N}
\left\lbrace
\frac{2}{\sqrt{2}an} \left(1 - e\right)^{\frac{1}{2}}A_1
+ 
\mathcal{O}\left[ \left(1-e\right)^{\frac{3}{2}} \right]
\right\rbrace
\]

\[
\ \ \ \ \ \ \ \
+ 
\mathcal{N}
\left\lbrace
-\frac{3}{\sqrt{2}an} \left(1 - e\right)^{\frac{1}{2}}A_2 
+ \mathcal{O}\left[ \left(1-e\right)^{\frac{3}{2}} \right]
\right\rbrace
f
\]

\begin{equation}
\ \ \ \ \ \ \ \ \ \ \
+ \mathcal{O}\left[f^2\right]
.
\end{equation}

\noindent{}

Finally, now we compare the above results with the evolution of the pericentre


\noindent{}

\[
\left( \frac{dq}{dt} \right)_{\rm out} \bigg|_{\begin{subarray}{c}{f \ {\rm about} \ 0^{\circ}} \\\epsilon \ {\rm about} \ 0 \end{subarray}}
= 0
\]

\begin{equation}
\ \ \ \ \ 
+ 
\mathcal{N}
\left\lbrace
\frac{1}{\sqrt{2}n} \left(1 - e\right)^{\frac{3}{2}}A_1 + \mathcal{O}\left[ \left(1-e\right)^{\frac{5}{2}} \right] 
\right\rbrace
f
+ \mathcal{O}\left[f^2\right]
.
\label{periout}
\end{equation}

\noindent{}

\noindent{}Equation (\ref{periout}) demonstrates that (i) the pericentre undergoes no evolution at all
in the leading order in true anomaly ($f^0$) term, (ii) in the next order term ($f^1$), the eccentricity
dependence is weaker [$\left(1-e\right)^{3/2}$] than that for any other orbital parameter, and (iii) this 
weaker leading-order term is independent of $A_2$ and $A_3$, so that the best way to change 
the pericentre is through radial perturbations. The first two points illustrate how robust 
the orbital pericentre is to changes compared to any other orbital element.

\section{Discussion}

Our conclusion is strong and independent of the detailed changes which a minor body or comet 
may experience after each pericentre passage if the lost volatiles do not constitute a 
large fraction of its total mass. In this case neither 
outgassing nor sublimation can perturb a 
WD exosystem minor planet or comet into the WD disruption sphere if 
the original orbit remains outside of that sphere.

Although potential time dependencies of many of the parameters that we held constant,
including $D_0$ and $v_{\rm g0}$, could certainly change a minor planet orbit, 
the actual pericentre cannot drift inward enough to transform an 
otherwise ``safe'' orbit into a ``disruptable'' one.  Consequently, the presence of planets
is necessary to perturb smaller bodies that reside within a few hundred au of the WD 
inside of the WD disruption sphere.
In main sequence exosystems, the invariance of the pericentre distance also holds as long as the 
extrasolar minor planet or comet
or is on a highly eccentric orbit.  These systems do not necessarily need to host currently
observable planets, as previous radial incursions from planet-planet scattering events can generate 
high eccentricities of smaller bodies that survive ejection 
\citep{verarm2005,verarm2006,rayetal2010,matetal2013}.
Other potential sources of such high eccentricities are stellar flybys and Galactic
tides; after a brief period of stellar mass loss, flybys and tides act in WD systems
in a similar manner to main sequence planetary systems, including the Solar system
\citep{veretal2014d}.

Because the timescales for orbital perturbations due to sublimation are short, and 
are often strongest at the first pericentre passage, they would likely dominate 
perturbations from WD radiation.
\cite{veretal2015b} demonstrated that the ever-dimming WD radiation will shrink 
and circularize the orbits of sub-metre-sized particles eventually due to 
Poynting-Robertson drag.  For the minor planets considered here,
which are km-sized, the dominant radiative effect would be the Yarkovsky effect \citep{veretal2015a}.
When activated, the Yarkovsky effect dominates Poynting-Robertson drag, but 
the timescale for Yarkovsky to act is still much longer than that from sublimation 
and GR.  Consequently, for volatile-rich comets, sublimation likely alters the 
orbits before Yarkovsky becomes important.  In these cases, sublimative forces 
in effect set the initial orbital conditions for other radiatively-induced 
perturbations.  Alternatively, for volatile-depleted minor planets, only the 
relative strength of GR and the Yarkovsky effect would be important.

A sublimative perturbation on the semimajor axis of even just a few hundredths
of an au may be significant.  In WD systems that contain planets, that difference 
can break or initiate a mean motion resonance with a planet.  Exo-Oort cloud 
comets which approach WDs \citep{alcetal1986,paralc1998,veretal2014c,stoetal2015} 
may be ejected into the interstellar medium.  In WD systems which are in the 
process of forming rings and discs from the tidal disruption of other 
minor planets, a change in semimajor axis will alter 
interactions with the resulting debris field during each pericentre passage.  
A significant debris field may
already exist anyway from the end of the giant branch phases of evolution due to YORP break-up of
km-sized asteroids, potentially extending to tens or hundreds of au \citep{veretal2014b}.

Finally, we note that the line between minor planets and comets may be fuzzy.  Even
in the Solar system, the distinction is not always easy to make \citep[see, e.g., the
Introduction of][]{jewetal2015}.  In fact, \cite{weilev1997} claimed that about 1 per
cent of the Solar system Oort cloud should comprise asteroids.  \cite{shaetal2015} 
more recently suggested that this population of asteroids should be about 
eight billion.

\section{Conclusions}

Hard observational evidence of volatile-rich circumstellar material 
in WD systems motivates studies about how active asteroids and comets 
self-perturb their orbits due to
the release of volatiles through sublimation and outgassing
during close pericentre passages with the star.
In this paper, we have derived the complete unaveraged equations of motion
(\ref{agen}-\ref{dfdt} and \ref{Astart}-\ref{Aend}) in orbital elements 
due only to sublimation, under some appropriate assumptions.  These equations are 
widely applicable to general extrasolar systems
assuming that planetary and external perturbations are negligible.
For Solar-system-like sublimation rates, the effect of GR must be 
included (Fig. \ref{set1}). We also considered the relative contribution
of outgassing in the transverse and nonplanar directions, and presented 
corresponding equations (\ref{subdadt}-\ref{subdqdt2})
that also may be applied to exosystems.

These results, when applied to polluted WD systems, ultimately reveal that 
major planets must exist in order to perturb active asteroids into the
WD disruption radius.  Without major planets, these highly-eccentric minor planets 
cannot change their pericentre distances significantly enough through the release of volatiles.  
Inspection alone of equations (\ref{qgen}), (\ref{qGR}), (\ref{dqdtP}) and (\ref{periout}) 
reveals why; the semimajor axis would change by many orders of magnitude more than the 
orbital pericentre in every case.

\section*{Acknowledgments}

We thank the referee for a probing and helpful report, and John H. Debes and Davide 
Farnocchia for useful discussions.
The research leading to these results has received funding from the European 
Research Council under the European Union's Seventh Framework Programme (FP/2007-2013) 
/ ERC Grant Agreement n. 320964 (WDTracer), EC Grant Agreement n. 282703 (NEOShield) and 
EC Grant Agreement n. 640351 (NEOShield-2) as well as Paris Observatory's ESTERS 
(Environement Spatial de la Terre : Recherche \& Surveillance) travel grants.

\appendix  
\onecolumn

\section{Sublimation with different power laws}

In order to assess how strongly dependent our sublimation findings are on the specific 
power-law prescription used in equation (\ref{nongrav2}), we now rederive the equations
of motion under more general considerations. Our physical motivation for the power-law 
exponent was observations from within the Solar system.  However, to allow for unknown
variations that can arise in the populations of small bodies in exosystems, we
assume the equation of motion for a sublimating small body is instead given by

\begin{equation}
\ddot{\vec{r}}_{\rm sub} \approx 
-\frac{G\left(M_{\rm WD} + M_{\rm c}\right)\vec{r}}{r^3}
-
\frac{\alpha}{M_{\rm c}}
\left(
\frac{r_{0}}{r}
\right)^{W}
\frac{\vec{r}}{r}
\label{nongrav3}
\end{equation}

\noindent{}

\noindent{}where $W$ is a positive integer and $\alpha$ is a constant.  
This integer requirement will allow
us to obtain averaged quantities analytically, and further bound the 
observationally-motivated value with $W = 2$ and $W = 3$.  This exercise 
will allow one to easily derive the equations of motion for an arbitrary
power-law perturbation to the two-body problem algebraically.

We find that the terms containing the $i$, $\Omega$ and the $C$ variables
remain unchanged such that

\begin{equation}
\left( \frac{da}{dt} \right)_{\rm sub}
=
\frac
{2 \alpha r_{0}^{W} \left(1 + e \cos{f} \right)^W}
{M_{\rm c} n a^W \left(1 - e^2 \right)^{W+\frac{1}{2}}}
\bigg[
C_1 \left(\cos{i}\cos{\Omega} - \cos{i}\sin{\Omega} + \sin{i} \right)
-
C_2 \left(\cos{\Omega} +\sin{\Omega} \right)
\bigg]
\label{Astart}
\end{equation}

\begin{equation}
\left( \frac{de}{dt} \right)_{\rm sub}
=
\frac
{\alpha r_{0}^{W}\left(1 + e \cos{f} \right)^{W-1} }
{2 M_{\rm c} n a^{W+1} \left(1 - e^2 \right)^{W-\frac{1}{2}}}
\bigg[
C_6 \left(\cos{i}\cos{\Omega} - \cos{i}\sin{\Omega} + \sin{i} \right)
 -
C_5 \left(\cos{\Omega} +\sin{\Omega} \right)
\bigg]
,
\end{equation}

\begin{equation}
\left( \frac{di}{dt} \right)_{\rm sub}
=
\frac
{\alpha r_{0}^{W}\left(1 + e \cos{f} \right)^{W-1} \cos{\left(f+\omega\right)} }
{M_{\rm c} n a^{W+1} \left(1 - e^2 \right)^{W-\frac{1}{2}}}
\sin{i} \left(\sin{\Omega} - \cos{\Omega} + \cot{i} \right)
,
\end{equation}

\begin{equation}
\left( \frac{d\Omega}{dt} \right)_{\rm sub}
=
\frac
{\alpha r_{0}^{W}\left(1 + e \cos{f} \right)^{W-1} \sin{\left(f+\omega\right)} }
{M_{\rm c} n a^{W+1} \left(1 - e^2 \right)^{W-\frac{1}{2}}}
\left(\sin{\Omega} - \cos{\Omega} + \cot{i} \right)
,
\end{equation}

\begin{equation}
\left( \frac{d\omega}{dt} \right)_{\rm sub}
=
\frac
{-\alpha r_{0}^{W}\left(1 + e \cos{f} \right)^{W-1} }
{2 M_{\rm c} n e a^{W+\frac{1}{2}} \left(1 - e^2 \right)^W}
\bigg[
C_8 \left(\cos{i}\cos{\Omega} - \cos{i}\sin{\Omega} \right)  + C_9\sin{i}
+
C_7 \left(\cos{\Omega} +\sin{\Omega} \right)
+
2 e \sin{\left(f+\omega\right)}\cos{i}\cot{i}
\bigg]
,
\label{Aend}
\end{equation}

\begin{equation}
\left( \frac{da}{dt} \right)_{\rm sub}^{\rm P}
=
\frac
{2 \alpha r_{0}^{W} \left(1 + e \cos{f} \right)^{W} }
{M_{\rm c} n a^{W} \left(1 - e^2\right)^{W+\frac{1}{2}}  }
\left(
C_{\rm 1P} - C_{\rm 2P} 
\right)
,
\label{dadtPW}
\end{equation}

\begin{equation}
\left( \frac{de}{dt} \right)_{\rm sub}^{\rm P}
=
\frac
{\alpha r_{0}^{W} \left(1 + e \cos{f} \right)^{W-1} }
{2 M_{\rm c} n a^{W+1} \left(1 - e^2\right)^{W-\frac{1}{2}}   }
\left(
C_{\rm 6P} - C_{\rm 5P} 
\right)
,
\label{dedtPW}
\end{equation}

\begin{equation}
\left( \frac{d\varpi}{dt} \right)_{\rm sub}^{\rm P}
=
-
\frac
{\alpha r_{0}^{W} \left(1 + e \cos{f} \right)^{W-1} }
{2 M_{\rm c} n e a^{W+\frac{1}{2}} \left(1 - e^2\right)^{W-\frac{1}{2}}   }
\left(
C_{\rm 7P} + C_{\rm 9P} 
\right)
,
\label{dcpidtPW}
\end{equation}

\noindent{}

Averaging the above equations presents difficulties analytically and
would typically require the use of computer algebra software.  Even
with that tool, we could obtain only the averaged planar equations
of motion.  We accomplish this task as in \cite{veras2014a}
by computing integrals symbolically where the integrand is a function
of an integral power of $\left(1 + e\cos{f}\right)$ multiplied by
single powers of $\sin{(kf)}$ and $\cos{(kf)}$, where $k$ is an integer
greater than or equal to 1.  

The resulting expressions, denoted by $\left\langle \right\rangle$, 
are long and we do not present them.
Instead, we expand the expressions in a Taylor series about
$\epsilon=1-e$, $\epsilon \ll 1$, and consider the leading term
in the expansion (denoted by a wide tilde). 
We hence report the lowest-order power
of $\left(1 - e\right)$ appearing in this expansion
in Table \ref{TablePower}.
 
Because this power is higher for the pericentre distance than for the
evolution of any other variable for $W \ge 2$, we conclude
that our main finding is robust for a wide variety
of steeper potential sublimation power-law profiles.

\begin{table}
 \centering
  \caption{Powers of $\left(1 - e\right)$ in the leading order term for
           the stated expressions expanded about high eccentricity.
           For $W \ge 2$, the pericentre distance is more resistant to change than
           any other variable.}
  \begin{tabular}{@{}ccccc@{}}
  \hline
 $W$ & $\widetilde{\left\langle \left( \frac{da}{dt} \right)_{\rm sub}^{\rm P} \right\rangle}$ & 
       $\widetilde{\left\langle \left( \frac{de}{dt} \right)_{\rm sub}^{\rm P} \right\rangle}$  &
       $\widetilde{\left\langle \left( \frac{dq}{dt} \right)_{\rm sub}^{\rm P} \right\rangle}$ &
       $\widetilde{\left\langle \left( \frac{d\varpi}{dt} \right)_{\rm sub}^{\rm P} \right\rangle}$ \\
 \hline
1    &   0    &   1/2   &   1/2    &   1/2   \\
2    &  -1    &    0    &   1/2    &    0    \\
3    &  -2    &   -1    &    0     &   -1    \\
4    &  -3    &   -2    &   -1     &   -2    \\
5    &  -4    &   -3    &   -2     &   -3    \\
6    &  -5    &   -4    &   -3     &   -4    \\
7    &  -6    &   -5    &   -4     &   -5    \\
\hline
\label{TablePower}
\end{tabular}
\end{table}

\twocolumn

\label{lastpage}
\end{document}